\documentclass{ws-ijmpc}
\usepackage[super]{cite} 
\usepackage{braket}
\usepackage{float}
\usepackage{graphicx} 
\usepackage{algorithm,float}
\usepackage[noend]{algpseudocode}

\begin{document}



\title{Evaluation of the Bilinear Condensate of the Planar Thirring Model in the Strongly Coupled Region}

\author{Jude Worthy}

\address{Department of Physics, College of Science, Swansea University\\
Swansea SA2 8PP, United Kingdom\\
jagworthy887140@swansea.ac.uk}

\author{Simon Hands}

\address{Department of Mathematical Sciences, University of Liverpool\\
Liverpool, L69 3BX, United Kingdom\\
simon.hands@liverpool.ac.uk}

\maketitle


\begin{abstract}
The planar Thirring model is thought to have a strongly coupled critical point for a single flavour of fermion. We look at the calculation of the bilinear condensate in this critical region, and its characterisation via an equation of state. Since the computation is numerically challenging we investigate improved Dirac operators. We present findings on different methods of calculation using a rational hybrid Monte Carlo scheme, and calculations of the bilinear condensate, an equation of state, and the associated critical exponents. Overlap and domain wall Dirac operators, and variants therein are considered.

\end{abstract}

\section{Introduction}

The Euclidean continuum formulation of the Thirring model \cite{2} for a single fermion field 
in $2+1d$ is given by \cite{1}:
\begin{equation}
S[\psi,\bar{\psi}]=\int d^3 x \bar{\psi}(\gamma_\mu \partial_\mu +m)\psi+\frac{g^2}{2}(\bar{\psi}\gamma_\mu\psi)^2
\label{eqn:thirring1}
\end{equation}
The self interacting term of the conserved current $\bar{\psi}\gamma_\mu\psi$
may be reformulated using  an auxiliary vector field $A_\mu$ to yield a
physcially equivalent action 
$S[\psi,\bar{\psi},A]=S_F[\psi,\bar{\psi},A]+S_G[A]$, where the fermionic action
$S_F$
takes the usual gauge invariant form while the auxiliary
action $S_G[A]$ is \it not \normalfont gauge invariant: 
\begin{equation}
S_F[\psi,\bar{\psi},A]=\int d^3 x \bar{\psi}(\gamma_\mu (\partial_\mu
+iA_\mu)+m)\psi \label{eqn:thirring2} \end{equation} \begin{equation}
\label{eqn:thirringaux} S_G[A]=\frac{1}{2g^2}\int d^3x A_\mu^2 
\end{equation} 
In
a planar (2+1$d$) formulation the Dirac field $\psi$ may be irreducible
(2-component), wherein the gamma matrices are given by the Pauli matrices, but
the Lagrangian (\ref{eqn:thirring1},\ref{eqn:thirring2}) is then not parity
invariant with the inclusion of a mass term $m\not=0$. Instead we choose a
reducible 4-component field, for which $m\bar\psi\psi$  is \normalfont parity
invariant. In particular there are two independent parity transforms, a
$\gamma_3$-transform, and a $\gamma_5$-transform, which under a reflection of
coordinates $x_\mu\to-x_\mu$ ($\mu=0,1,2$) transform with, for $\gamma_j \in
\{\gamma_3,\gamma_5\}$: 
\begin{equation}\label{eqn:paritysymmetry} \begin{split}
\psi\to i\gamma_j\psi \; \;  & ; \; \; \bar{\psi} \rightarrow
-i\bar{\psi}\gamma_j \\ A\to-A \; \;  & \\ \end{split} 
\end{equation} 
The
4-component field also admits a $U(2)$ global symmetry described by eqns.
(\ref{eqn:u2symmetry}), explicitly breaking to a $U(1)\otimes U(1)$ with a
non-zero mass; eqns. \ref{eq:seq3}, \ref{eq:seq4} do not hold for $m\not=0$.
These rotations are analogous to chiral symmetry in 3+1$d$ formulations, and 
in the presence of interactions may
also be spontaneously broken at and beyond a critical point. With fixed
coordinates  
\begin{subequations}\label{eqn:u2symmetry} \begin{align}
\psi\rightarrow e^{i\alpha}\psi \; \; & ; \; \; \bar{\psi} \rightarrow
\bar{\psi}e^{-i\alpha} \\ \psi\rightarrow e^{i\alpha\gamma_3\gamma_5}\psi \; \;
& ; \; \; \bar{\psi} \rightarrow \bar{\psi}e^{-i\alpha\gamma_3\gamma_5} \\
\psi\to e^{i\alpha\gamma_3\psi} \; \; & ; \; \; \bar{\psi} \rightarrow
\bar{\psi}e^{i\alpha\gamma_3} \label{eq:seq3} \\ \psi\rightarrow
e^{i\alpha\gamma_5}\psi \; \; & ; \; \; \bar{\psi} \rightarrow
\bar{\psi}e^{i\alpha\gamma_5} \label{eq:seq4} \end{align} 
\end{subequations} 
We
may separate the Dirac spinors into ``left handed'' ($+$)  and ``right handed''
($-$)
components. In 3+1$d$ this distinction is uniquely determined by the chirality operator
$\gamma_5$, with projectors $P^5_\pm=(1\pm\gamma_5)/2$ isolating each
handedness. In 2+1$d$ the distinction is not unique and we may also choose
$\gamma_3$, with $P^3_\pm=(1\pm\gamma_3)/2$. A $\gamma_3$-parity transform swaps
a $P^3$-left spinor into a $P^3$-right spinor and vice versa, and similarly for
the $\gamma_5$-parity transform. 

We also have a continuous family of $U(2)$-equivalent mass terms \cite{3},
from which we extract and label $m\bar{\psi}\psi$, $im\bar{\psi}\gamma_3\psi$ as
the standard and twisted mass formulations respectively.

On a space-time lattice with spacing $a$ we encounter the Nielsen-Ninomiya no-go theorem, rendering $U(2)$
symmetry incompatible with the locality of the Dirac operator, and no lattice
doublers. A work-around is given by the Ginsparg-Wilson relations \cite{4}, in
which the $U(2)$ transformations are adjusted according to\cite{3}:
\begin{equation}\label{eqn:GWtrans} \begin{split} \psi\to
e^{i\alpha\gamma_3(1-\frac{aD}{2})}\psi \; \;  ; \; \; \bar{\psi} \rightarrow
\bar{\psi}e^{i\alpha\gamma_3(1-\frac{aD}{2})} \\ \psi\rightarrow
e^{i\alpha\gamma_5(1-\frac{aD}{2})}\psi \; \; ; \; \; \bar{\psi} \rightarrow
\bar{\psi}e^{i\alpha\gamma_5(1-\frac{aD}{2})}, \\ \end{split} 
\end{equation} 
where the Dirac operator $D$ satisfies
\begin{equation}
\{\gamma_3,D\}=2aD\gamma_3D;\;\;
\{\gamma_5,D\}=2aD\gamma_5D;\;\;
[\gamma_3\gamma_5,D]=0.
\end{equation}
For
any such $D$, in the continuum limit $a\to0$ we see that $U(2)\to
U(1)\otimes U(1)$ is recovered. This is contingent on the locality of the
operator, which ensures $aD\to 0$. 

The continuum limit corresponds to a critical point in the phase diagram of the
lattice Thirring model. Besides explicit symmetry breaking the Thirring model has
been demonstrated to spontaneously break $U(2)\to U(1)\otimes U(1)$, analagously to chiral
symmetry breaking, with the bilinear condensate $\braket{\bar{\psi}\psi}$ as the
order parameter. Around this critial point we construct an empirical equation of state
\cite{5}, defining the behaviour of the order parameter in the critical region.
\begin{equation}\label{eqn:EoS}
m=A(\beta-\beta_c)\braket{\bar{\psi}\psi}^{\delta
-1/\beta_m}+B\braket{\bar{\psi}\psi}^\delta. \end{equation}

The existence and calculated properties of the critical point have been found to
depend on the type of lattice model used. The critical flavour number is
the maximum number of fermion flavours which allow a critical point. Domain wall
techniques \cite{1} find this critical flavour number to be $N_f=1$, distinct
from both staggered \cite{7} ($N_f>1$) and SLAC \cite{8} ($N_f<1$) lattice fermion
formulations. The values of the associated critical exponents $\beta_m,\delta$ differ as well.

We continue to choose the overlap Dirac operator \cite{9,10} 
\begin{equation}\label{eqn:dolA} D_{OL}
=\frac{1+im\gamma_3}{2}+\gamma_3\text{sgn}(H)\frac{1-im\gamma_3}{2}
\\ 
\end{equation} 
as most suitable for our discretisation of the fermionic terms, since it obeys
the Ginsparg-Wilson relations \cite{4}, enabling the recovery of the desired
global U(2) symmetry in the continuum limit.  Choices in the implementation of
the overlap operator include the kernel of the sign function $H$, the
approximation to the sign function, a further regularization parameter $M$, and
the form of the bare mass term \cite{3,11}, all of which we expect to be
physically equivalent in the U(2)-symmetric continuum limit. After an
approximation to the sign function is chosen the operator is often referred to
as a truncated overlap operator.  The domain wall operator $D_{DW}$, specified
more fully in Section~\ref{sec:DWF} below, is expressed with an additional
non-physical dimension $x_3$ separating two open boundaries or ``domain walls''
by an extent $L_s$, may be chosen to be exactly related to a truncated overlap
operator \cite{11,12,13} and hence an overlap operator in the limit
$L_s\to\infty$. In such cases the regularization parameter $M$ is equivalently
the domain wall height. Consequently, formulations may be chosen so that
numerically identical results are found with either $D_{OL}$ or $D_{DW}$, as
shown in Section~\ref{sec:DWOLrel}. The domain wall formulation is more
convenient for the generation and evolution of the auxiliary fields via the
Rational Hybrid Monte Carlo (RHMC) method.  When the Wilson kernel ($H=\gamma_3
D_W$) 
is chosen it is preferable to use the overlap operator given by
(\ref{eqn:dolA}) in the measurement calculations, where $D_W[A]$ is the usual
Wilson Dirac operator specified  in (\ref{eqn:wilsondiracop}) below. When the Shamir
kernel ($H=\gamma_3 D_W(2+D_W)^{-1}$) is chosen the domain wall formulation is
generally preferable.  Significant cost benefits in condensate calculations have
been demonstrated with the twisted mass form \cite{3} as specified in
(\ref{eqn:dolA}). 

Using domain wall Dirac operators $D_{DW}$ a critical point has been found with a single
fermion flavour but not with two or more, and hence we use RHMC
\cite{14} for the generation of the auxiliary fields - the single
fermion in the effective action is given by $(D_{DW}^\dagger D_{DW})^{-1/2}$ for which we
use a rational approximation.

The lattice Thirring model for a single fermion is given by action\footnote{Note
that the power $1/2 \to -1/2$ when moving from the Grassmann field $\psi$ in the
action, to the complex-valued pseudofermion $\psi$ in the effective action.}
\begin{equation}\label{eqn:thirringaction} S_{\text{Th}}=\psi^\dagger
(D_{DW}[U_\mu]^\dagger D_{DW}[U_\mu])^{1/2} \psi + \beta A_\mu^2 
\end{equation} 
where the coupling parameter
$\beta=\frac{a}{g^2}$ is adimensional. The link field $U$ in the operator $D_{DW}[U]$ 
is generated with non-compact relation $U_\mu=1+iaA_\mu$ rather than the
compact relation $U_\mu=e^{iaA_\mu}$ used for lattice QED. Periodic boundary conditions are
applied in the two spatial dimensions and require no adjustment, but
an anti-periodic boundary condition in the temporal dimension is specified for the
time components of the link field on the time dimension boundary, so that 
$U_0(T,x,y)=-U_0(1,x,y)$ for all $x,y$. 
The lattice spacing $a$ is set to 1 in all of the following.

\section{Condensate Measurements}

The bilinear condensate order parameter is defined by
\begin{equation}
{{\partial\ln{\cal Z}}\over{\partial m}}=\left\langle{\rm Tr}\left[{{\partial\ln
D(m)}\over{\partial m}}\right]\right\rangle=\left\langle {\rm Tr}\left[ D^{-1}{{\partial D}\over{\partial
m}}\right]\right\rangle,
\end{equation}
where the partition function is defined by the functional integral 
${\cal Z}=\int{\cal D}\psi{\cal  D}\bar\psi{\cal  D}Ae^{-S[\psi,\bar\psi,A]}$.
We consider Dirac operators which separate out the mass component
$D[U]=D^0[U]+mD^M[U]$. We estimate the bilinear condensate over $N$ instances
with distinct auxiliary fields $U_i$, using Monte Carlo integration
\begin{equation}\label{eqn:blcond}
\braket{\bar{\psi}\psi}=\frac{1}{N}\sum_{i=1}^N \frac{1}{V}\text{Tr} [D^M[U_i]
D[U_i]^{-1}] 
\end{equation} 
where $V$ is the number of vertices in the lattice.
We use the RHMC method\cite{14,15} outlined
further below, for the generation of auxiliary fields $A_i$, and consequently
$U_i$, for each instance. We choose overlap Dirac operators (see (\ref{eqn:dol})
below),
so that $D_{OL}^0[U] = \frac{1}{2}+\frac{1}{2}V[U]$ and we distinguish two mass
variants, the standard $D_{OL}^{M1}[U] = \frac{1}{2}-\frac{1}{2}V[U]$, and
twisted $D_{OL}^{M3}[U] = \frac{i\gamma_3}{2}-V[U]\frac{i\gamma_3}{2}$, in which
$V[U] = \gamma_3 \text{sgn}[H[U]]$. More details are given below, but regardless
of the choice of kernel $H[U]$ we have the condensate instances \begin{equation}
\begin{split} \hat
O_{OL}^1=\frac{1}{V}\text{Tr}\left[\frac{1}{1-m}((D^1_{OL})^{-1}-1)\right]; \\ \hat
O_{OL}^3=\frac{1}{V}\text{Tr}\left[\frac{-1}{i\gamma_3+m}((D^3_{OL})^{-1}-1)\right] \\
\end{split} \end{equation} 
corresponding to $D_{OL}^1=D_{OL}^0+mD_{OL}^{M1}$ and
$D_{OL}^3=D_{OL}^0+mD_{OL}^{M3}$. A noisy estimator may be used to calculate the
trace, according to algorithm \ref{alg:noisyestimator}, with $N_N=10$ used in
this work.

\begin{algorithm}[H]
\caption{Noisy estimator for the trace of the inverse of a complex matrix $M$}\label{alg:noisyestimator}
\begin{algorithmic}[1]
\Function{CalculateNoisyEstimator}{$M$}
\State Set $s=0$
\For {$n=1,N_N$}  \% $N_N$ is the number of noisy estimators per instance
  \State $\eta=\mathcal{CN}(0,1)$ \% vector of complex Gaussian random numbers
  \State $\xi=M^{-1}\eta$
  \State $s=s+\eta^\dagger \xi$
\EndFor
\State \textbf{return} $s/N_N$
\EndFunction
\end{algorithmic}
\end{algorithm}

\section{Dirac Operators}

We want measurements calculated with the overlap operator. Domain wall operators
may be formally equivalent to overlap operators when the Pauli-Villars 
terms to compensate for the influence of unphysical
bulk fields with $1<x_3<L_s$ are
included (see below), 
and it is often computationally advantageous to use
the domain wall formulation. Variants of $D_{OL}$, and similarly the
equivalent $D_{DW}$, include the choice of mass term - standard
or twisted, the choice of kernel - Shamir or Wilson, and the choice of kernel
approximation method - hyperbolic tangent (HT) or Zolotarev (Z). Whether to use
$D_{DW}$ or $D_{OL}$ formulation is context
dependent, although table \ref{tab:DiracGuide} provides a guide in which we
distinguish between the valence fermion Dirac operator - that used in the Monte
Carlo integration of eqn. (\ref{eqn:blcond}) - and the sea fermion operator - that
used in the effective action (\ref{eqn:effectiveaction}) below, used in the hybrid
Monte Carlo algorithms for the generation of the auxiliary fields.
\begin{table}[ht] \tbl{Guide to choice of overlap formulation}
{\begin{tabular}{@{}ccc@{}} \toprule \hline & Valence  & Sea   \\ \hline
Strongly coupled  &  Twisted Wilson Zolo $\Leftrightarrow D_{OL}$  & Twisted
Wilson HT $\Leftrightarrow D_{DW}$  \\
Weakly coupled &Twisted Shamir HT $\Leftrightarrow D_{DW}$  & Twisted Shamir HT
$\Leftrightarrow D_{DW}$  \\
\hline \end{tabular} \label{tab:DiracGuide}} 
\end{table} 
All of these variants
are constructed from the Wilson Dirac operator $D_W[U]$. Derivatives of the Dirac
operators are required in hybrid Monte Carlo algorithms, and may similarly be
constructed via the derivatives of $D_W$. The operators and
algorithms for their application are given below.

\subsection{Wilson Dirac Operator}

The continuum Euclidean Dirac operator is given by
$D_E\equiv\gamma_\mu(\partial_\mu+iA_\mu)+m$, and its conjugate is
$D_E^\dagger\equiv-\gamma_\mu(\partial_\mu+iA_\mu)+m$. We choose complex link
fields given by $U_\mu=1+iA_\mu$ rather than the usual compact
$U_\mu=e^{iA_\mu}$ form, eliminating six-point and higher order interaction coupling terms in the
discretisation of (\ref{eqn:thirring1}) once the auxiliary fields are integrated
over. The lattice Wilson Dirac operator $D_W$ applied to
lattice node $n$ is then given by  
\begin{equation} \label{eqn:wilsondiracop}
\begin{split} (D_W \psi)(n) & = \frac{1}{2}\sum_{\mu=0,1,2} \bigl[\gamma_\mu
(U_\mu(n)\psi(n+\hat{\mu})-U_{\mu}(n-\hat{\mu})^\dagger\psi(n-\hat{\mu})) \\ &
-(U_\mu(n)\psi(n+\hat{\mu})-2\psi(n)+U_{\mu}(n-\hat{\mu})^\dagger\psi(n-\hat{\mu}))\bigr]
\\ & + m\psi(n) \end{split} 
\end{equation} 
The first line of
(\ref{eqn:wilsondiracop}) gives the Dirac term, corresponding to the massless part
of the continuum operator, and the second line of (\ref{eqn:wilsondiracop})
is the Wilson term, introduced to remedy the so-called lattice doubling problem
\cite{22}. The conjugate operator $D_W^\dagger$ changes the sign of the Dirac
term, but not the Wilson term. The derivative operator is given by
\begin{equation} \label{eqn:wilsondiracopderivs} \left(\frac{\partial D_W}{\partial
A_{x,\mu}} \psi\right)(n) = \begin{cases} \frac{1}{2} \gamma_\mu \psi(x+\hat{\mu})i -
\frac{1}{2} \psi(x+\hat{\mu})i, & n=x \\ \frac{1}{2} \gamma_\mu \psi(x)i
+\frac{1}{2} \psi(x)i, & n=x+\hat{\mu} \\ 0 & \text{otherwise} \\ \end{cases}
\end{equation} 
and hence  for arbitrary complex spinor fields $\eta,\nu$
\begin{equation}\label{eqn:wilsonforcederivs}
\begin{split} \eta^\dagger \frac{\partial D_W}{\partial A_{x,\mu}}\nu & =
\frac{1}{2}\eta(x)^\dagger\gamma_\mu i \nu(x+\hat{\mu}) -\frac{1}{2}
\eta(x)^\dagger i\nu(x+\hat{\mu}) \\ &  +\frac{1}{2} \eta(x+\hat{\mu})^\dagger i
\gamma_\mu \nu(x) + \frac{1}{2} \eta(x+\hat{\mu})^\dagger i\nu(x)   \\
\end{split} 
\end{equation} 
Again, the conjugate $\frac{\partial
D_W^\dagger}{\partial A_{x,\mu}}$ changes the sign of the Dirac term. We can
then calculate the Wilson derivative matrix, $F_{D_W}$, and conjugate
$F^\dagger_{D_W}$ used in the RHMC algorithm as set out in algorithm
\ref{alg:WilsonDerivatives}, where 
\begin{equation}\label{eqn:wilsonforce}
F_{D_W}=\eta^\dagger \frac{\partial D_W}{\partial A} \nu \;\;\; ; \;\;\;
F^\dagger_{D_W}=\eta^\dagger \frac{\partial D_W^\dagger}{\partial A}\nu
\end{equation}
 
\begin{algorithm}[H] \caption{Wilson Dirac
Derivatives}\label{alg:WilsonDerivatives} \begin{algorithmic}[1]
\Function{WilsonDerivs}{$\eta$,$\nu$,DAG} \% eqn \ref{eqn:wilsonforce} via eqn
\ref{eqn:wilsonforcederivs} \ForAll{$x,\mu$} \If {DAG=FALSE} \State
$F((x,\mu)=\eta(x)^\dagger\gamma_\mu i \nu(x+\hat{\mu})/2 +
\eta(x+\hat{\mu})^\dagger\gamma_\mu i \nu(x)/2$ \State $- \eta(x)^\dagger i
\nu(x+\hat{\mu})/2 + \eta(x+\hat{\mu})^\dagger\gamma_\mu i \nu(x)/2$ \ElsIf
{DAG=TRUE} \State $F((x,\mu)=-\eta(x)^\dagger\gamma_\mu i \nu(x+\hat{\mu})/2 +
\eta(x+\hat{\mu})^\dagger\gamma_\mu i \nu(x)/2$ \State $-\eta(x)^\dagger i
\nu(x+\hat{\mu}/2 + \eta(x+\hat{\mu})^\dagger\gamma_\mu i \nu(x)/2$ \EndIf
\State \textbf{return} $F$ \EndFor \EndFunction \end{algorithmic}
\end{algorithm}

\subsection{Overlap Dirac Operator}

The overlap Dirac operator \cite{9,10} is usually given by $D^1_{OL}
=\frac{1+m}{2}+\frac{1-m}{2}V$. The twisted mass formulation enabled by the
planar dimensionality may be expressed with the following in which
$V=\gamma_3\text{sgn}(H)$\cite{12} (Cf. eqn.~(\ref{eqn:dolA})).  
\begin{equation}\label{eqn:dol}
\begin{split} D^3_{OL} & =\frac{1+im\gamma_3}{2}+V\frac{1-im\gamma_3}{2} \\
D_{OL}^{3\dagger} & =\frac{1-im\gamma_3}{2}+\frac{1+im\gamma_3}{2}V^\dagger \\
\end{split} 
\end{equation} 
$H\equiv\gamma_3 D_W$ specifies the Wilson kernel. 
We are also interested the formulation with Shamir kernel for which
$H=\frac{\gamma_3 D_W}{2+D_W}$ and which corresponds to the usual domain wall
formulation due to  Shamir~\cite{24}. 
Noting $\text{sgn}(H)\equiv H(H^2)^{-1/2}$, and $\gamma_3 D_W \gamma_3 = D_W^\dagger$,
for the Wilson kernel we have 
\begin{equation}\label{eqn:vol} 
\begin{split} 
V & =D_W(D_W^\dagger D_W)^{-1/2} \\ 
V^\dagger & =D_W^\dagger(D_W^\dagger D_W)^{-1/2} \\ 
\end{split}
\end{equation} 
Note that $m$ is the bare mass; the Wilson Dirac operator (\ref{eqn:wilsondiracop}) in the
kernel $H$ has mass term $M\bar\psi\psi$ with a regularisation parameter $-2<M<0$ 
in place of the usual mass term $m\bar\psi\psi$.
$M=-1$ in this work. The inverse square root, exponent $p=-1/2$, must be
approximated, and we do so with a partial fraction formulation of a
corresponding rational function. This is then sometimes referred to as a
truncated overlap operator. We will also want to evaluate other operators
similarly with powers other than $p=-1/2$, notably $p=1/4$. For arbitrary
operators $D$ and different powers $p$, the partial fraction formulation of a
rational function approximation of $(D^\dagger D)^p$ is given by
\begin{equation}\label{eqn:diracratfuncs} (D^\dagger D)^p = \alpha_{0,p}+\sum_i
\frac{\alpha_{i,p}}{D^\dagger D+\beta_{i,p}} 
\end{equation} 
The denominators may be evaluated, for example, successively with a conjugate gradient algorithm, or
simultaneously with a multishift conjugate gradient algorithm~\cite{19}. 
When the Shamir kernel is used this is slow to calculate and it
is computationally faster to calculate measurements using a domain wall
formulation instead; this formulation does provides a cross check of any code
however.

The choice of coefficients $\alpha,\beta$ is significant. From the hyperbolic tangent
approximation to the sign function, $\text{sgn}(x)\approx
\text{tanh}(n\text{tanh}^{-1} x)$, we have the partial fraction expression for
even $n$\cite{11}: 
\begin{equation} \text{sgn}(x)\approx
\frac{2x}{n}\sum_{j=0}^{n/2-1}\frac{1+(\text{tan}\frac{(j+1/2)\pi}{n})^2}{x^2+(\text{tan}\frac{(j+1/2)\pi}{n})^2}
\end{equation} 
The inverse root power may be taken from
$x^{-1/2}=\text{sgn}(x^{1/2})/x^{1/2}$. Alternatively we may use an optimal
Zolotarev approximation\cite{26} to the sign function\footnote{The Zolotarev
approximation to the inverse root power is, of course, similarly related}. The
coefficients depend on the specified range of the approximation. That is, we
have the Zolotarev functions $\mathcal{Z}_{n}^{x_L,x_R}(x)\approx \text{sgn}(x)$
in the range $x\in[x_L,x_R]$, and with accuracy increasing with order $n$. This
range must encompass the spectral range of the kernel $H$. These can be
calculated analytically\cite{28,29}, or more generally using the Remez
algorithm, which also allows the calculation of coefficients for other powers.
There is freely available software to calculate the coefficients via the
iterative Remez algorithm \cite{23}. There should be {\it a priori\/} knowledge of the
spectral range if one is to avoid recalculating the coefficients for every
distinct auxiliary field.

\subsection{Domain Wall Operators}
\label{sec:DWF}

The domain wall operators have an extra dimension of discrete extent $L_s$, and
we label an extended fermion field $ \Psi$, denoting slices indexed along  the extra
dimension $j=1,\ldots,L_s$ with $\psi_j$. Application of the domain wall operator may be carried
out by application of the Wilson Dirac operator on different slices according to
the following \cite{20,12,24}. We will specify both Shamir and Wilson kernel
variants, each with a standard and twisted mass formulation.

\subsubsection{Shamir Kernel Domain Wall Dirac Operator}

We consider two forms: $D_{SDW}^{3}=D_{SDW}^0+mD_{SDW}^{M3}$, and
$D_{SDW}^{1}=D_{SDW}^0+mD_{SDW}^{M1}$. In either case the massless expression
$\Psi^\prime=D^0_{SDW}\Psi$ has components
\begin{equation}\label{eqn:domwallshamir0} \psi^\prime_j = \begin{cases}
(D_W+I)\psi_j  - P_-\psi_{j+1} & j=1 \\
  - P_+\psi_{j-1} + (D_W+I) \psi_j  - P_-\psi_{j+1}  & 1 < j < L_s \\
  - P_+\psi_{j-1} + (D_W+I) \psi_j  & j = L_s \\ \end{cases} 
\end{equation} 
with projectors $P_\pm\equiv P^3_\pm$ defined earlier.
The conjugate operator, with  $\bar \Psi^\prime = (D_{SDW}^0)^\dagger \Psi$ is
given by 
\begin{equation}\label{eqn:domwallshamir0dag} \bar\psi^\prime_j =
\begin{cases} (D_W^\dagger+I)\psi_j  - P_+ \psi_{j+1} & j=1 \\
  - P_-\psi_{j-1} + (D_W+I) \psi_j  - P_+\psi_{j+1}  & 1 < j < L_s \\
  - P_-\psi_{j-1} + (D_W+I) \psi_j  & j = L_s \\ \end{cases} 
\end{equation}
Standard ($mD_{SDW}^{M1}$) mass components and their conjugate counterparts
contribute 
\begin{equation}\label{eqn:domwallsgamirmass1} \begin{split}
\psi^\prime_1 =  mP^+ \psi_{L_s} \;\; & \;\; \bar\psi^\prime_1 =  mP^-
\psi_{L_s} \\ \psi^\prime_{L_s} =  mP^- \psi_1 \;\; & \;\; \bar\psi^\prime_{L_s}
= mP^+ \psi_1\\ \end{split} 
\end{equation} 
The twisted ($mD_{SDW}^{M3}$) mass
components contribute\footnote{Note $P^+\gamma_3=\gamma_3P^+=P^+$ and
$P^-\gamma_3=\gamma_3P^-=-P^-$} 
\begin{equation}\label{eqn:domwallshamirmass3}
\begin{split} \psi^\prime_1 =  -im P^+ \gamma_3 \psi_{L_s} \;\; & \;\;
\bar\psi^\prime_1 =  im\gamma_3 P^- \psi_{L_s} \\ \psi^\prime_{L_s} = -im P^-
\gamma_3 \psi_1 \;\; & \;\; \bar\psi^\prime_{L_s} = im\gamma_3  P^+\psi_1 \\
\end{split} 
\end{equation} 
The derivatives $\partial D_{SDW}/\partial A$ are described along with the Wilson kernel
in \ref{sec:WDW} below, since they are identical to the diagonal components of that
scheme.

\subsubsection{Wilson Kernel Domain Wall Dirac Operator}
\label{sec:WDW}

We again distinguish two forms via the standard and twisted mass terms,
$D_{WDW}^{3}=D_{WDW}^0+mD_{WDW}^{M3}$, and
$D_{WDW}^{1}=D_{WDW}^0+mD_{WDW}^{M1}$. We now further denote $D_j^+=\omega_j D_W
+I $, $D_j^-=\omega_j D_W - I $. Setting the coefficients $\omega_j=1$
corresponds to the  hyperbolic tangent approximation of the sign function. To
obtain the Zolotarev approximation we set $\omega_j=1/u_j$ where $u_j$
are the roots of $\mathcal{Z}_n^{x_L,x_R}(x)-1=0$. Code to calculate these roots
and the Zolotarev coefficients directly is also freely available\cite{30}. For
the massless component we have 
\begin{equation}\label{eqn:domwallwilson0}
\psi^\prime_j = \begin{cases} D_j^+\psi_j  + D_j^- P_-\psi_{j+1} & j=1 \\ D_j^-
P_+\psi_{j-1} + D_j^+ \psi_j  + D_j^- P_-\psi_{j+1}  & 1 < j < L_s \\ D_j^-
P_+\psi_{j-1} + D_j^+ \psi_j  & j = L_s \\ \end{cases} 
\end{equation} 
The
conjugate operator, with  $\bar\psi^\prime_j \equiv ((D_{WDW}^0)^\dagger) \bm
\Psi)(j)$, $D_j^{\dagger}+=\omega_j D_W^\dagger +I $, $D_j^{\dagger-}=\omega_j
D_W^\dagger - I $, is 
\begin{equation}\label{eqn:domwallwilson0dag}
\bar\psi^\prime_j = \begin{cases} D_j^{\dagger+}\psi_j  + P_+
D_{j+1}^{\dagger-}\psi_{j+1} & j=1 \\ P_- D_{j-1}^{\dagger-} \psi_{j-1} +
D_j^{\dagger+} \psi_j  + P_+ D_{j+1}^{\dagger-} \psi_{j+1}  & 1 < j < L_s \\ P_-
D_{j-1}^{\dagger-} \psi_{j-1} + D_j^{\dagger+} \psi_j  & j = L_s \\ \end{cases}
\end{equation}
Standard ($mD_{WDW}^{M1}$) mass components and their conjugate
counterparts contribute 
\begin{equation}\label{eqn:domwallwilsonmass1}
\begin{split} \psi^\prime_1 =  -mD_1^-P^+ \psi_{L_s} \;\; & \;\;
\bar\psi^\prime_1 =  -mP^- D_{L_s}^{\dagger-} \psi_{L_s} \\ \psi^\prime_{L_s} =
-mD_{L_s}^- P^- \psi_1 \;\; & \;\; \bar\psi^\prime_{L_s} = -mP^+
D_{1}^{\dagger-} \psi_1\\ \end{split} 
\end{equation} 
The twisted
($D_{WDW}^{M3}$) mass components contribute
\begin{equation}\label{eqn:domwallwilsonmass3} \begin{split} \psi^\prime_1 =
-imD_1^-P^+ \gamma_3 \psi_{L_s} \;\; & \;\; \bar\psi^\prime_1 =  im\gamma_3 P^-
D_{L_s}^{\dagger-} \psi_{L_s} \\ \psi^\prime_{L_s} = -imD_{L_s}^- P^- \gamma_3
\psi_1 \;\; & \;\; \bar\psi^\prime_{L_s} = im\gamma_3 P^+ D_{1}^{\dagger-}
\psi_1\\ \end{split} 
\end{equation} 
The derivative term $\frac{\partial
D_{DW}}{A_{x,d}}\Psi$ has component slices $\psi^\prime_j$ for the massless
parts given by 
\begin{equation}\label{eqn:domwallderivs0} \psi^\prime_j =
\left(\frac{\partial D_{DW}^0}{\partial A_{x,d}}\Psi\right)(j) = \begin{cases} \omega_j
\frac{\partial D_W}{\partial A_{x,\nu}}\psi_j  + \omega_j \frac{\partial
D_W}{\partial A_{x,\nu}} P_-\psi_{j+1} & j=1 \\ \omega_j \frac{\partial
D_W}{\partial A_{x,\nu}} P_+\psi_{j-1} + \omega_j \frac{\partial D_W}{\partial
A_{x,\nu}} \psi_j  + \omega_j \frac{\partial D_W}{\partial A_{x,\nu}}
P_-\psi_{j+1}  & 1 < j < L_s \\ \omega_j \frac{\partial D_W}{\partial
A_{x,\nu}}P_+ \psi_{j-1} + \omega_j \frac{\partial D_W}{\partial A_{x,\nu}}
\psi_j  & j = L_s \\ \end{cases} 
\end{equation}
\begin{equation}\label{eqn:domwallderivs0dag} \bar \psi^\prime_j =
\left(\frac{\partial D_{DW}^{0\dagger}}{\partial A_{x,d}}\Psi\right)(j) = \begin{cases}
\omega_j \frac{\partial D_W^\dagger}{\partial A_{x,d}}\psi_j  + P_+ \omega_{j+1}
\frac{\partial D_W^\dagger}{\partial A_{x,d}}\psi_{j+1} & j=1 \\ P_-
\omega_{j-1} \frac{\partial D_W^\dagger}{\partial A_{x,d}}\psi_{j-1} +\omega_j
\frac{\partial D_W^\dagger}{\partial A_{x,d}}\psi_j  + P_+ \omega_{j+1}
\frac{\partial D_W^\dagger}{\partial A_{x,d}}\psi_{j+1}  & 1 < j < L_s \\ P_-
\omega_{j-1} \frac{\partial D_W^\dagger}{\partial A_{x,d}}\psi_{j-1} +\omega_j
\frac{\partial D_W^\dagger}{\partial A_{x,d}}\psi_j  & j = L_s \\ \end{cases}
\end{equation} 
and for the mass parts by
\begin{equation}\label{eqn:domwallderivsmass} \psi^\prime_j =\frac{\partial
D_{DW}^{M}}{\partial A_{x,d}}\psi = \begin{cases} -\omega_1 \frac{\partial
D_W}{\partial A_{x,d}} P^+ c \psi_{L_s} & j= 1 \\ -\omega_{L_s} \frac{\partial
D_W}{\partial A_{x,d}} P^- c \psi_1 & j=L_s \\ 0 & \text{otherwise} \end{cases}
\end{equation} 
\begin{equation}\label{eqn:domwallderivsmassdag} \bar
\psi^\prime_j =\frac{\partial D_{DW}^{M\dagger}}{\partial A_{x,d}}\psi =
\begin{cases} -\omega_{L_s} c^\dagger P^- \frac{\partial D_W^\dagger}{\partial
A_{x,d}} \psi_{L_s} & j= 1 \\ -\omega_1 c^\dagger P^+ \frac{\partial
D_W^\dagger}{\partial A_{x,d}} \psi_1 & j=L_s \\ 0 & \text{otherwise}
\end{cases} \end{equation} 
For the standard mass term $c=c^\dagger=1$, and for
the twisted mass term $c=i\gamma_3$, $c^\dagger=-i\gamma^3$.

Algorithms \ref{alg:domainwallderivs} and \ref{alg:domainwallmassderivs} put together the domain wall derivative matrix, $F_{D_{DW}}=\bm{\eta}\frac{\partial D_{DW}}{\partial A}\bm{\nu}$, and conjugate $F^\dagger_{D_{DW}}=\bm{\eta}\frac{\partial D^\dagger_{DW}}{\partial A}\bm{\nu}$.
\begin{algorithm}[H]
\caption{DomainWallDerivs. Returns array $F_{D_{DW}}=\bm{\eta}\frac{\partial D_{DW}}{\partial A}\bm{\nu}$ or $F_{D_{DW}}^\dagger=\bm{\eta}\frac{\partial D_{DW}^\dagger}{\partial A}\bm{\nu}$ via eqns. \ref{eqn:domwallderivs0},\ref{eqn:domwallderivs0dag}.}\label{alg:domainwallderivs}
\begin{algorithmic}[1]
\Function{DomWallDerivs}{$\bm{\eta}$,$\bm{\nu}$,MTYPE,DAG}
\State $F=0$
\State \% diagonal components
\For {$j=1,...,L_s$}   
\State $F = F+\Call{WilsonDerivs}{\bm{\eta}(j),\bm{\nu}(j),DAG}$
\EndFor
\State \% if using Shamir kernel, return here \\
\State \% upper and lower diagonal components
\For {$j=1,...,L_s-1$} 
\State $F = F+\Call{WilsonDerivs}{\bm{\eta}(j),\bm{\nu}(j+1),DAG}$
\State $F = F+\Call{WilsonDerivs}{\bm{\eta}(j+1),\bm{\nu}(j),DAG}$
\EndFor 

\State \% Mass terms 
\State $F = F + \Call{DomWallMassDerivs}{\bm{\eta},\bm{\nu},MTYPE,DAG}$
\State \textbf{return} $F$

\EndFunction
\end{algorithmic}
\end{algorithm}

\begin{algorithm}[H]
\caption{DomainWallMassDerivs. Returns the mass components of $F_{D_{DW}}$ via eqns. \ref{eqn:domwallderivsmass}, \ref{eqn:domwallderivsmassdag}.}\label{alg:domainwallmassderivs},
\begin{algorithmic}[1]
\Function{DomWallMassDerivs}{$\bm{\eta}$,$\bm{\nu}$,MTYPE,DAG}
\State \% mass components
\If{(MTYPE.eq.STANDARD).AND.(DAG.eq.FALSE)}
\State $l_1=\bm{\eta}(1)$  ;  $r_1=-\omega_1P^+\bm{\nu}(L_s)$
\State $l_2=\bm{\eta}(L_s)$  ;  $r_2=-\omega_{L_s}P^-\bm{\nu}(1)$
\ElsIf{(MTYPE.eq.STANDARD).AND.(DAG.eq.TRUE)}
\State $l_1=-\omega_{L_s}P^-\bm{\eta}(1)$  ;  $r_1=\bm{\nu}(L_s)$
\State $l_2=-\omega_{L_s}P^+\bm{\eta}(L_s)$  ;  $r_2=\bm{\nu}(1)$
\ElsIf{(MTYPE.eq.TWISTED).AND.(DAG.eq.FALSE)}
\State $l_1=\bm{\eta}(1)$  ;  $r_1=-i\omega_1P^+\gamma_3\bm{\nu}(L_s)$
\State $l_2=\bm{\eta}(L_s)$  ;  $r_2=-i\omega_{L_s}P^-\gamma_3\bm{\nu}(1)$
\ElsIf{(MTYPE.eq.TWISTED).AND.(DAG.eq.TRUE)}
\State $l_1=-i\omega_{L_s}\gamma_3 P^-\bm{\eta}(1)$  ;  $r_1=\bm{\nu}(L_s)$
\State $l_2=-i\omega_{L_s}\gamma_3 P^+\bm{\eta}(L_s)$  ; $r_2=\bm{\nu}(1)$
\EndIf
\State \textbf{return} $\Call{WilsonDerivs}{l_1,r_1} + \Call{WilsonDerivs}{l_2,r_2}$
\EndFunction
\end{algorithmic}
\end{algorithm}

\section{Generation of Auxiliary Fields - Hybrid Monte Carlo} \label{sec:HMC}

We want to generate independent random auxiliary fields $A$ with probability
distribution $P[A]\propto e^{-S[A]}$, for our Thirring action $S[A]$, eqn.
\ref{eqn:thirringaction}. We generate a sequence $A^0,...,A^N $ using a hybrid
Monte Carlo method. For $N$ large enough, $A_N$ will take the specified
distribution $\propto e^{-S[A]}$, regardless of the choice of $A_0$. Each step
of the sequence comprises a hamiltonian dynamics component, which generates a
proposed field $A^\prime$, followed by a Monte Carlo acceptance step, in which
the proposed field is either accepted or rejected as the next field in the
sequence, as set out in algorithm \ref{alg:hmc}.  
\begin{algorithm}
\caption{Hybrid Monte Carlo generation of a randomly distributed auxiliary
field. Step 1: Generate a proposed auxiliary field. Step 2: Monte Carlo
acceptance}\label{alg:hmc}
\begin{algorithmic}[1] \Function{generateNextAux}{$A^0$} \For{s=1 to S} \State
[$A^\prime,P^{s-1},P^\prime,\phi]=$\Call{genPropAux}{$A^{s-1}$} \State $A^s
=$\Call{MCacceptance}{$A^{s-1}$,$A^\prime$,$P^{s-1}$,$P^\prime$,$\phi$} \EndFor
\State \textbf{return} $A^S$ \EndFunction \end{algorithmic} 
\end{algorithm}

\subsection{Hamiltonian Dynamics}

The proposed field $A^\prime$ is found by time marching the artificially created
hamiltonian system $H=\frac{1}{2}P^2+S[A]$ with dynamical equations
\begin{equation} \begin{split} \dot A & = \frac{\partial H}{\partial P} = P \\
\dot P & = -\frac{\partial H}{\partial A} = -\frac{\partial S[A]}{\partial A} \\
\end{split} \end{equation} 
Since the Hamiltonian expression has no explicit time
dependence, the total energy $h=\sum H_{i,\mu}$ summed over contributions from
all sites and links of the  lattice is constant; hence under perfect
evolution of the equations the system remains on the surface of constant $h$. A leap-frog
method is then used to march to the solution, with the momentum terms in $P$ taking a
half step at the beginning and end of the trajectory, the length of which is
randomised around a fixed average time scale. The algorithm\cite{11} is set out
in the pseudo code \ref{alg:hts}.

\begin{algorithm}
\caption{Hamiltonian Dynamics Step for general Dirac operator $D$. Time-march the auxiliary field according to the artificially created hamiltonian system.}\label{alg:hts}
\begin{algorithmic}[1]
\Function{genPropAux}{$A^0$}
\State Set $P^0\propto \mathcal{N}(0,1)$
\State Set $r \propto \mathcal{CN}(0,1)$
\State Set $\phi=D^\dagger r $ for effective action $S_{eff}=\phi^\dagger (D^\dagger D)^{-1} \phi$
\State Set timestep size $\triangle t$ and average number of timesteps $T_{av}$.
\State $\text{dSdA} =\Call{Force}{A^0}$ 
\State $P^{1/2}=P^0+\triangle t/2 \;\text{dSdA}$
\For{$t=0,...,T_{max}$}
  \State $A^{t+1}=A^t+\triangle t P^{t+1/2}$
  \State $\text{dSdA}=\Call{Force}{A^{t+1}}$
  \State Set $y\propto \mathcal{U}(0,1)$
  \If {$y>1/T_{av}$}
      \State $P^{t+3/2}=P^{t+1/2}-\triangle t \; \text{dSdA}$
  \Else
      \State $P^{t+3/2}=P^{t+1/2}-\triangle t/2 \; \text{dSdA}$
      \State goto 100 
  \EndIf
\EndFor
\State $P^{T_{max}}=P^{T_{max}-1/2}+\triangle t/2 \; \text{dSdA}$
\State 100 continue
\State \textbf{return} $A^\prime=A^t$,$P^0$,$P^\prime=P^t$,$\phi$
\EndFunction
\end{algorithmic}
\end{algorithm}
Our Thirring action splits into fermionic and auxiliary field components
$S[\psi,\bar\psi,A]=S_F[\psi,\bar\psi,A]+S_G[A]$. The fermionic component uses Grassmann-valued fields
$\psi,\bar\psi$,
so we
must use an equivalent action with \it effective \normalfont fermionic component
$S[\phi,\phi^\dagger,A]=S_{eff}[\phi,\phi^\dagger,A]+S_G[A]$, 
such that integration $\int {\cal D}\phi{\cal D}\phi^\dagger e^{-S_{eff}}$
over the complex-valued $\phi,\phi^\dagger$ produces 
an identical result to the Grassmann integral 
$\int{\cal D}\psi{\cal D}\bar\psi e^{-S_F}$,
facilitating the calculation of the \it force
\normalfont term $\frac{\partial S[A]}{\partial A}$ incorporated in the
Hamiltonian time marching. In order to generate auxiliary fields for
measurements with a single fermion with operator $D$, we have
$S_{eff}[\phi,\phi^\dagger,A]=\phi^\dagger (D^\dagger D)^{-1/2} \phi$, in which the pseudofermion
field $\phi$ is complex. This formulation \cite{14} specifies the rational
hybrid Monte Carlo method (RHMC). 

We use domain wall operators defining $D\equiv
(D^{M1}_{DW}(1))^{-1}D^{M3}_{DW}(m)$ since as $L_s\to\infty$ these have the same determinant as
the overlap operator $D_{OL}$~\cite{12}. After some rearrangement via the determinant, and denoting
$MS\equiv D_{DW}^{M1}(1)$ and $MT\equiv D_{DW}^{M3}(m)$, we use\cite{18}:
\begin{equation}\label{eqn:effectiveaction} S_{eff}[A] =\phi^\dagger [(MS)^\dagger
MS]^{1/4} [(MT)^\dagger MT]^{-1/2} [(MS)^\dagger MS]^{1/4}\phi \\
\end{equation} 
In this context degrees of freedom governed by $MS$ are sometimes referred to as
Pauli-Villars fields, but it is computationally much more efficient to have
$MS,MT$ act on the same pseudofermion field $\phi$.
In the domain wall formulation $\phi$ may be considered an
extradimensional pseudofermion, or a set of pseudofermions. Either way in line 4
of algorithm \ref{alg:hts}, the effective action
(\ref{eqn:effectiveaction}) implies $D\equiv ((MT)^\dagger MT)^{1/4}((MS)^\dagger
MS)^{-1/4}$. Then $\phi =((MS)^\dagger MS)^{-1/4} ((MT)^\dagger MT)^{1/4} r$ is set
at the beginning of each trajectory and remains fixed until the next trajectory.
The complex random gaussian vector $r\propto\mathcal{CN}(0,1)$ is created independently
on each site of the $2+1+1d$ lattice.

As per the action, the force terms split into a fermionic component and
an auxiliary component given simply by 
$\frac{\delta S_G}{\delta A_{x,\mu}}=2 \beta A_{x,\mu}$.
Force terms of the fermionic component, that is the derivatives of the fermionic
effective action, are given by
\begin{eqnarray}
\frac{\delta S_{eff}}{\delta A_{x,\mu}} &=& 2\text{Re}
\left[\phi^\dagger \frac{\delta((MS)^\dagger MS)^{1/4}}{\delta A_{x,\mu}}
\bar{\phi} \right]
+ \hat{\phi}^\dagger \frac{\delta((MT)^\dagger MT)^{-1/2}}{\delta
A_{x,\mu}}\hat{\phi} \label{eq:domforceA} \\ 
\hat{\phi} &=& ((MS)^\dagger MS)^{1/4} \phi \label{eq:domforcecoeffsA} \\
\bar{\phi} &=& ((MT)^\dagger MT)^{-1/2} \hat{\phi} \label{eq:domforcecoeffsB}
\end{eqnarray} 
Noting for any matrix $\frac{\partial}{\partial x} M^{-1}= - M^{-1}
\frac{\partial M}{\partial x} M^{-1}$, and using a partial fraction expansion of
the power terms, we get 
\begin{equation}\label{eq:domforceB} \begin{split}
\frac{\delta S_{eff}}{\delta A_{x,\mu}} = & -2\text{Re} \left[\phi^\dagger \Bigl[\sum_i
\alpha_{i,1/4} ((MS)^\dagger MS+\beta_{i,1/4})^{-1} \frac{\delta  (MS)^\dagger
MS}{\delta A_{x,\mu}}  ((MS)^\dagger MS+\beta_{i,1/4})^{-1}\Bigr] \bar{\phi} \right] \\
- & \hat{\phi}^\dagger
\left[\sum_i \alpha_{i,-1/2} ((MT)^\dagger MT+\beta_{i,-1/2})^{-1} \frac{\delta
(MT)^\dagger MT}{\delta A_{x,\mu}}  ((MT)^\dagger MT+\beta_{i,-1/2})^{-1}\right]
\hat{\phi} \\ \end{split} 
\end{equation} 
Since $\frac{\delta  (MS)^\dagger
MS}{\delta A_{x,\mu}}=(MS)^\dagger \frac{\delta MS}{\delta A_{x,\mu}}+\frac{\delta
(MS)^\dagger}{\delta A_{x,\mu}}MS$ ($\not=2\text{Re}[(MS)^\dagger
\frac{\delta MS}{\delta A_{x,\mu}}]$)\footnote{$M+M^\dagger=2\text{Re} [M]$ if
$M$ is hermitian. $MS$ and $MT$ are not hermitian, but all $M^\dagger M$ are
hermitian.}, we have the force array $F=\frac{\partial S}{\partial A}$ given in
algorithm \ref{alg:fermrhmcforce}. 

\begin{algorithm}[H]
\caption{FermRHMCForce. $\text{RATIONAL}(M,\phi,p)$ is a function returning $(M^\dagger M)^p\phi$. $\text{IMdagMpC}(M,c,\phi)$ is a function returning $(M^\dagger M+c)^{-1}\phi$.}\label{alg:fermrhmcforce}
\begin{algorithmic}[1]
\Function{RHMCForce}{$\phi$} \% follows eqn. \ref{eq:domforceB} with terms $\frac{\delta  MS^\dagger MS}{\delta A_{x,\mu}}$ expanded.
\State $F=0$
\State $\hat \phi = \Call{Rational}{MS,\phi,pow=1/4}$ \% eqn. \ref{eq:domforcecoeffsA}
\State $\bar \phi = \Call{Rational}{MT,\hat\phi,pow=-1/2}$ \% eqn. \ref{eq:domforcecoeffsB}
\ForAll {$i\in\beta_{i,1/4}$} \% First line of eqn. \ref{eq:domforceB}
\State \% calc sl,sr,vl,vr
\State $s_l=\Call{IMdagMpC}{MS,\beta_{i,1/4},\phi}$
\State $s_r=\Call{IMdagMpC}{MS,\beta_{i,1/4},\bar\phi}$
\State $v_l=MS s_l$
\State $v_r=MS s_r$
\State $v_1=\Call{DomWallDerivs}{s_l^\dagger, v_r, STANDARD, DAG=TRUE}$
\State $v_2=\Call{DomWallDerivs}{v_l^\dagger, s_r, STANDARD, DAG=FALSE}$
\State $F=F-2\alpha_{i,1/4}\text{Re}[v_1+v_2]$
\EndFor
\ForAll {$i\in\beta_{i,-1/2}$} \% Second line of eqn. \ref{eq:domforceB}
\State \% calc t,w
\State $t=\Call{IMdagMpC}{MT,\beta_{i,-1/2},\hat \phi}$
\State $w=MT t$
\State $v_1=\Call{DomWallDerivs}{t^\dagger, w, TWISTED,DAG=TRUE}$
\State $v_2=\Call{DomWallDerivs}{w^\dagger, t, TWISTED,DAG=FALSE}$
\State $F=F-\alpha_{i,-1/2}(v_1+v_2)$
\EndFor
\State \% Apply anti-peridic bcs in time dimension
\ForAll {$x,y$}
\State $F(x,y,T)=-F(x,y,1)$  
\EndFor
\State \textbf{return} $F$
\EndFunction
\end{algorithmic}
\end{algorithm}

Also note $\frac{\delta MS}{\delta A_{x,\mu}}\equiv\frac{\delta MT}{\delta
A_{x,\mu}}$ when the Shamir kernel is used, but not when the Wilson kernel is
used. Pedagogical implementations of the methods used in this paper can be found
online.\cite{31} 

\subsection{Monte Carlo Acceptance}

The proposed auxiliary field $A^\prime$ is accepted as the next field in the
sequence with probability $a=\text{min}\{1,\text{exp}(h^{prev}-h^\prime)\}$
where the total energy (Hamiltonian), $h$, is given by \begin{equation}
\begin{split} h & =h_g+h_f+h_p \\ h_g =\sum_{x,\mu} \frac{\beta}{2} A_{x,\mu}^2
\;\; ; \;\;  & h_f = S_{eff}[\phi,\phi^\dagger,A] \;\; ; \;\; h_p
=\sum_{x,\mu} \frac{1}{2} P_{x,\mu}^2 \\ \end{split} 
\end{equation} 
Pseudocode is
given in algorithm \ref{alg:mca}.  \begin{algorithm}[H] \caption{Monte Carlo
Acceptance}\label{alg:mca} \begin{algorithmic}[1]
\Function{MCaccept}{$A^{prev}$,$A^\prime$,$P^{prev}$,$P^\prime$} \State
$h^{prev}=$\Call{Hamiltonian}{$A^{prev}$,$P^{prev}$} \State
$h^\prime=$\Call{Hamiltonian}{$A^\prime$,$P^\prime$} \State r is U(0,1) \State
$a=\text{min}\{1,\text{exp}(-h^{prev}+h^\prime)\}$ \If {$r < a$} \State
\textbf{return} $A^\prime$ \Else \State \textbf{return} $A^{prev}$ \EndIf
\EndFunction \end{algorithmic} \end{algorithm}

\subsection{Relation between Overlap and Domain Wall Dirac Operators}\label{sec:DWOLrel}

Early in the development of the overlap operator it was noted \cite{10} that the
domain wall operator\cite{21} could be developed from the overlap formalism.
There is an operator, $K_{DW}$, which also has an extra dimension and can be
defined via the domain wall operators, for which $(K_{DW}\Psi)(1)=D_{OL}\psi$
identically when $\Psi=\{\psi,0,\cdots,0\}$ is a vector of slices in the extra
dimension. We may also calculate the inverse of the overlap operator indirectly
with $D_{OL}^{-1}\psi=(K_{DW}^{-1}\Psi)(1)$. For the Wilson kernel with twisted
mass it is expressed \cite{12,13} \begin{equation}\label{eqn:kdwm1}
K_{DW}=C^\dagger (D^1_{DW}(1))^{-1}D^3_{DW}(m)C \end{equation} in which the
``compacting'' operators $C$ and $C^\dagger$ are given by
\begin{equation}\label{eqn:cmatrix} \begin{split} (C\Psi)(j) = & \begin{cases}
P_- \psi_j  +  P_+ \psi_{j+1} & 1 \le j < L_s \\ P_- \psi_j  +  P_+
\psi_{j+1-L_s} & j = L_s \\ \end{cases} \\ (C^\dagger \Psi)(j) = (C^{-1} \psi)_j
& \begin{cases} P_- \psi_j  +  P_+ \psi_{j-1+L_s} & 1 \\ P_- \psi_j  +  P_+
\psi_{j-1} & 1 < j \le L_s \\ \end{cases} \\ \end{split} 
\end{equation} 
It is
important to note that for the Pauli-Villars terms, $(D^1_{DW}(1))^{-1}$ in
(\ref{eqn:kdwm1}), the twisted mass formulation is {\em not}  used. 

\section{Results}

Following algorithm \ref{alg:hmc} to generate sequences of auxiliary fields, and
algorithm \ref{alg:noisyestimator} to measure instances of the condensate we
find the following results on a $12^2\times12$ lattice, and look at some further
properties of the algorithms.

\subsection{Acceptance Rates of the Monte Carlo Step}

\begin{figure}[H]
\begin{center}
\vspace{-0.5cm}
\includegraphics[scale=0.22]{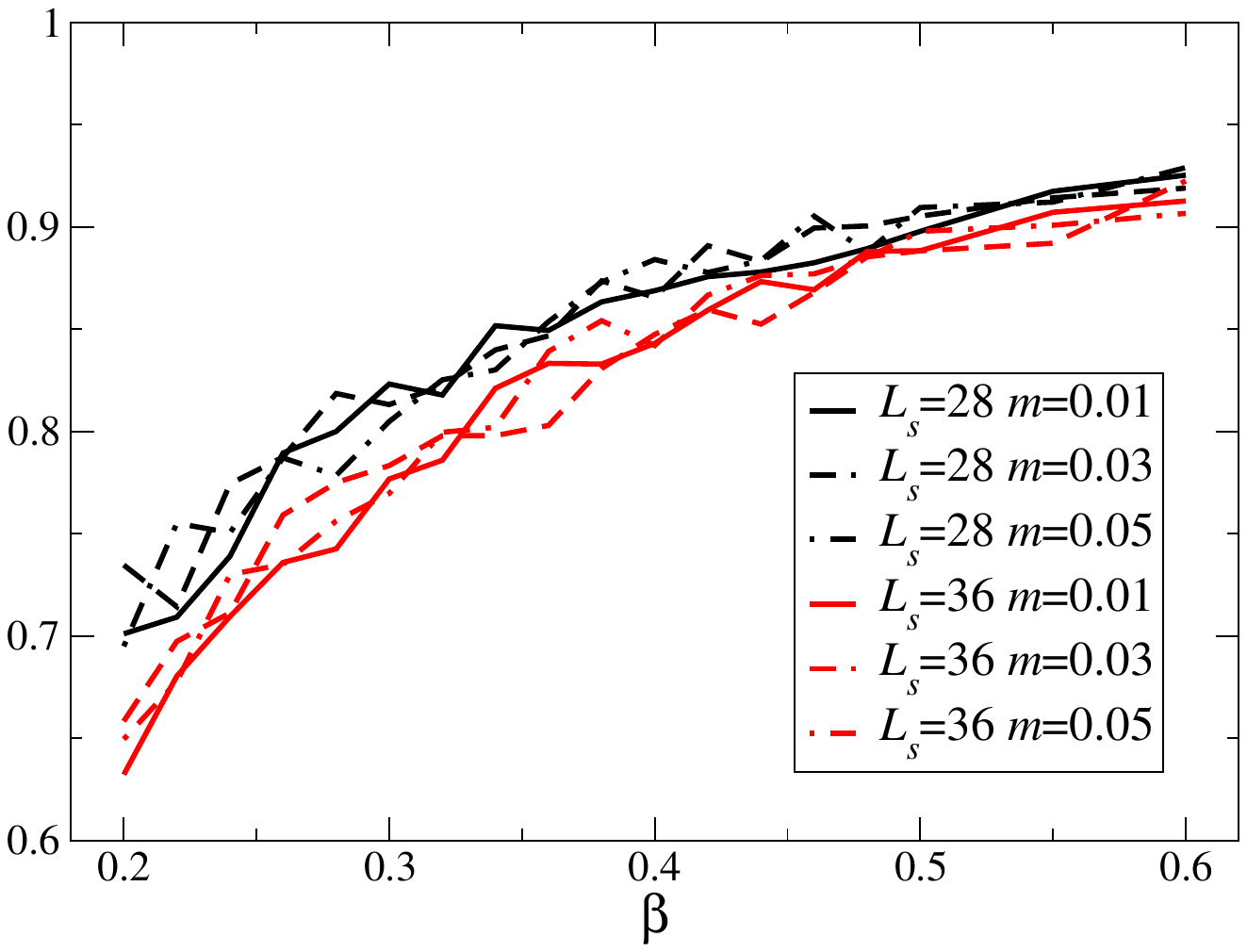}
\includegraphics[scale=0.22]{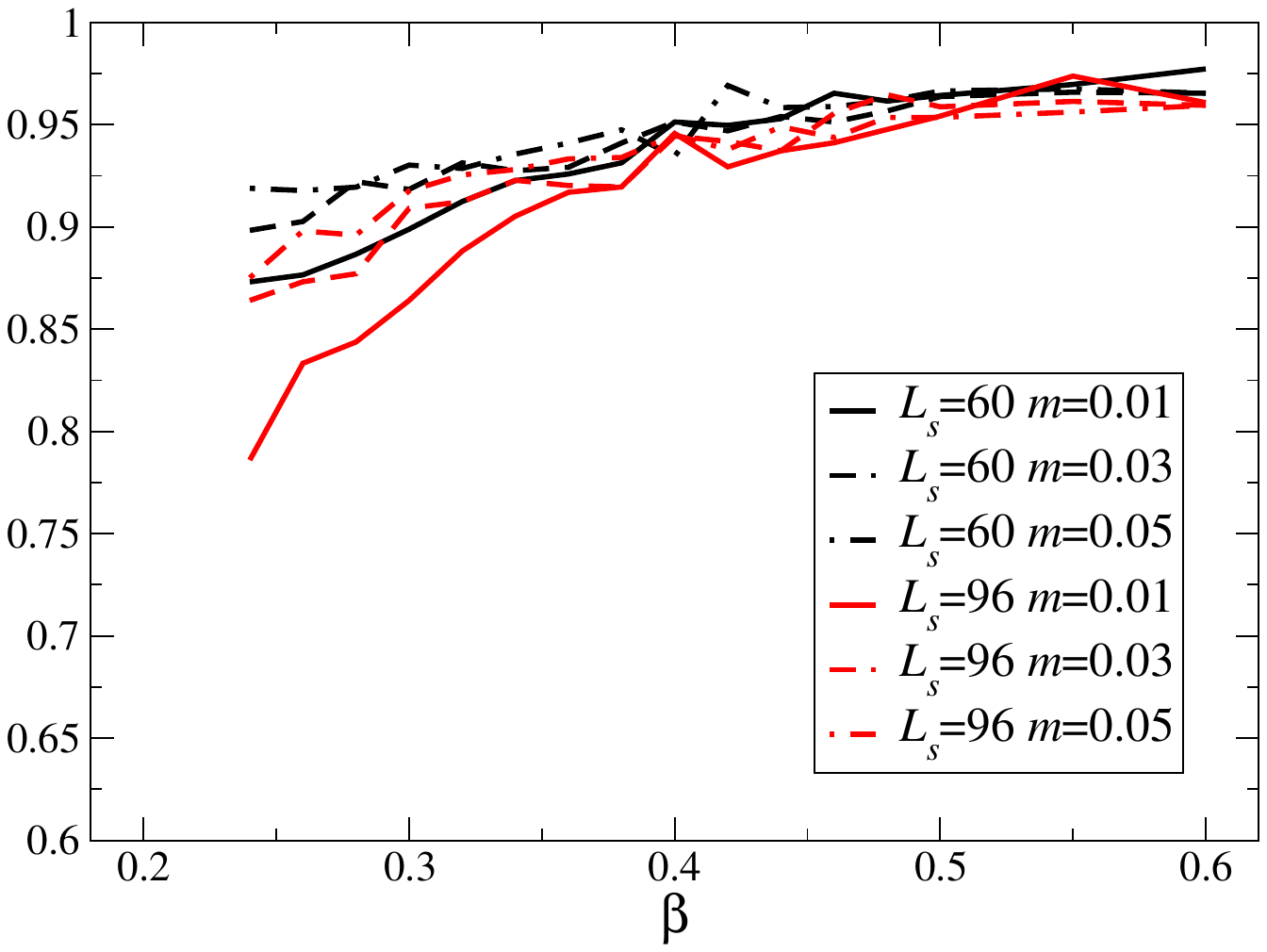}
\vspace{-0.5cm}
\end{center}
\caption{Acceptance Rate in the Monte Carlo Step. $L_s$ is the extent of the
extra dimension in the domain wall formulation, and spacetime lattice is
$12^3$. Note the critical region is estimated  to be 0.33-0.34. Left panel:
Wilson kernel. Right panel: Shamir kernel.} \label{fig:acceptance} \end{figure}

The RHMC method takes an extant auxiliary field, and then finds a proposed
auxiliary field by marching it forward using a fictitious hamiltonian dynamics
step, moving it through its phase space. The proposed field is then accepted or
not accepted in the Monte Carlo step. We look at the acceptance rates of this
Monte Carlo step in fig. \ref{fig:acceptance}, varying the non-dimensional
coupling strength $\beta$. The average trajectory length is 0.5 with a time step
$\triangle t=0.05$ for both the Wilson and Shamir cases. 

It is clear that under these conditions the Wilson formulation has a lower
acceptance rate. In the weaker coupling range, the acceptance rate is between
0.9 and 0.95 for the Wilson formulation and between 0.95 and 1 for the Shamir
formulation, and dropping to 0.65 to 0.75 and 0.8 to 0.95 respectively at the
stronger coupling end of the plots. 

It is notable that the acceptance rate decreases with extra-dimensional extent
$L_s$ since it is to be expected that the acceptance rate stops decreasing after
achieving sufficiently large $L_s$.\footnote{There appears to be no $L_s$ volumetric effect
on the hamiltonian dynamics step, $\dot P = -\frac{\partial S[A]}{\partial A}$; 
that is the effective action has little if any $L_s$-dependence for large enough $L_s$, and does
not lead to a destabilisingly large effective time step.} 
We will see below that using the Wilson kernel seems to have lower $L_s$
requirements than the Shamir kernel in the critical region, 
supporting with the hypothesis
that for fixed $L_s$ the Wilson scheme is closer to the correct U(2)-invariant dynamics. 

\subsection{Eigenvalue Extrema and Condition Number of Kernel}

\begin{figure}[H] \begin{center}
\vspace{-0.5cm}
\includegraphics[scale=0.22]{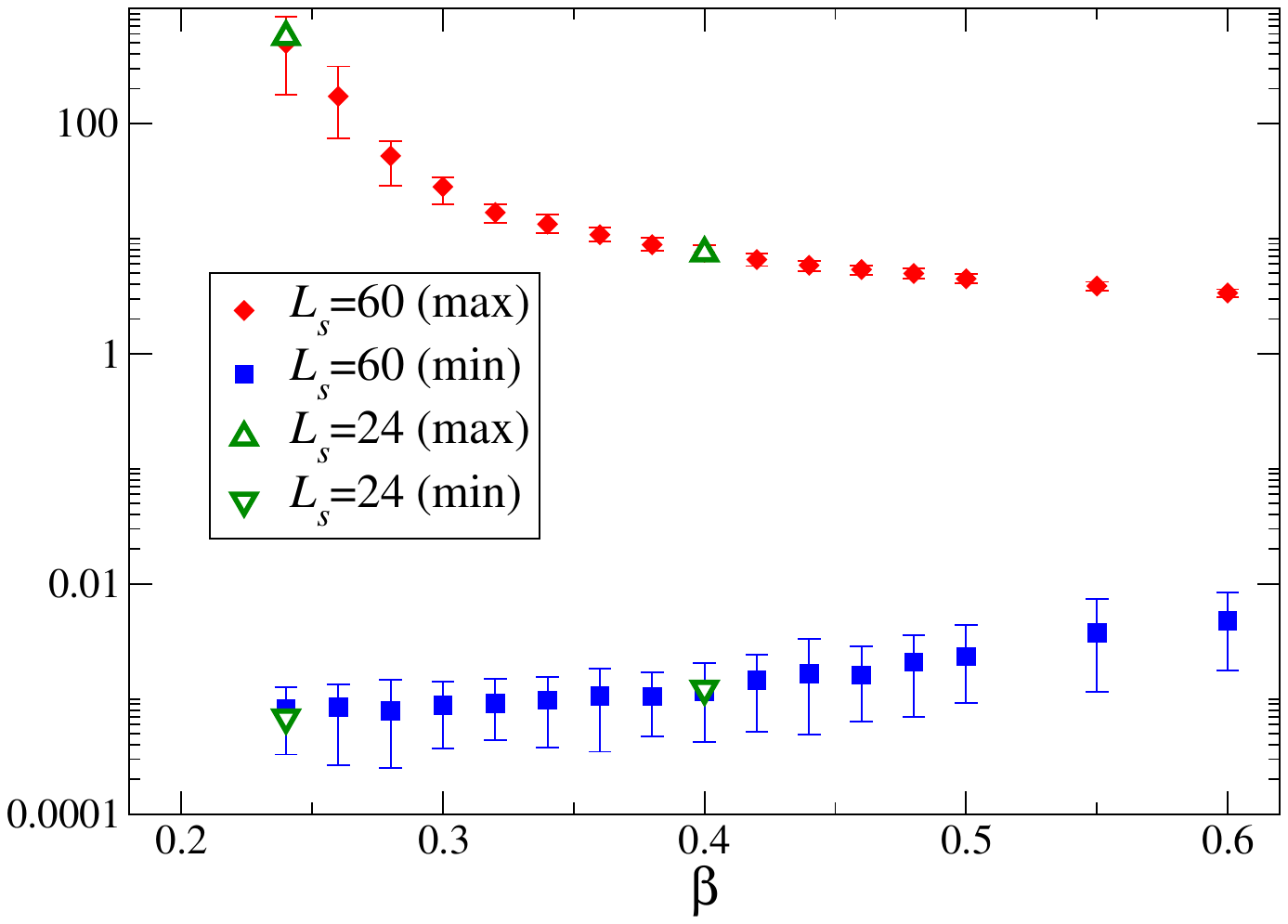}
\includegraphics[scale=0.22]{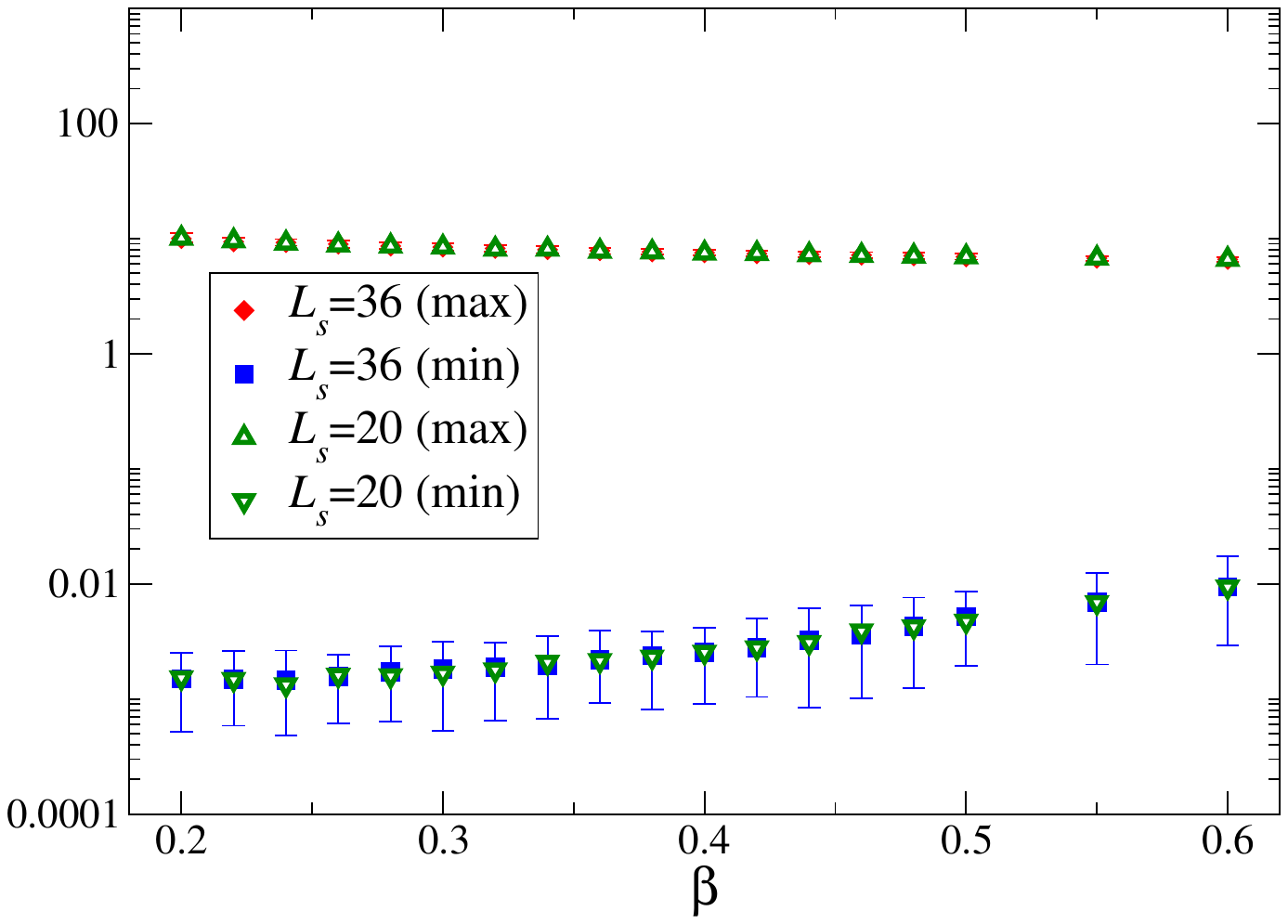}
\vspace{-0.5cm}
\end{center} \caption{Min/max kernel
eigenvalues obtained with $m=0.01$. Left panel: Shamir kernel. Right panel: Wilson kernel.
Error bars show the first and ninth deciles of the eigenvalue data across the
ensemble.}
\label{fig:KernelEigs} 
\end{figure}

It is useful to look at the eigenvalue range of the kernels. Since the auxiliary
fields are generated dynamically, they are technically dependent on the $L_s$
extent of the domain wall. When viewed from the overlap perspective this is not
considered an extra physical dimension, but merely an expression of the level of
accuracy of approximation. The left panel of Figure \ref{fig:KernelEigs} shows
the maximum and minimum eigenvalues for the Shamir kernel, and we see that the
average values are largely independent of either the mass $m$ or of $L_s$.
At least 100 configurations  were used for each data point. Auxiliary fields for the
evaluation of the Shamir kernel were generated with the Shamir kernel domain
wall operator, and the fields for the evaluation of the Wilson kernel were
generated using the Wilson kernel domain wall operator. Unlike formulations using
compact link fields, the eigenvalues are not bounded from above and, for the
Shamir kernel, there is a significant increase in the maximum eigenvalue beyond
the critical $\beta_c\approx0.33$.\cite{16} 
Plots for the Wilson kernel are shown in the
right panel of Figure \ref{fig:KernelEigs}. 
These eigenvalues provide a
guide for choosing the Zolotarev range to be used in the overlap operator,
unless the range is to be reset for every auxiliary configuration. Since the
latter is costly, especially for the dynamic step, it is generally preferable to
choose a fixed range. Although the eigenvalues are strictly only bounded below
by zero, a practical range can be identified from the plots. 
We have found failing to adhere strictly to the upper bound causes a larger
changes 
in the value  of $\langle\bar\psi\psi\rangle$ than varying the lower bound, 
which can be ascribed to a finite operator renormalisation on variation of
the UV cutoff.

The $L_s$-independence, at least for all $L_s$ above an unexplored lower
bound, combined with an a priori belief that the condensate measurements 
require significantly higher $L_s$ values, suggests the possibility of using
different $L_s$ values for the sea-fermions (the Dirac operator used in the
generation of the auxiliary fields) and the valence fermions (the Dirac operator
used for the condensate measurements). 

\begin{figure}[H]
\begin{center} 
\vspace{-0.5cm}
\includegraphics[scale=0.22]{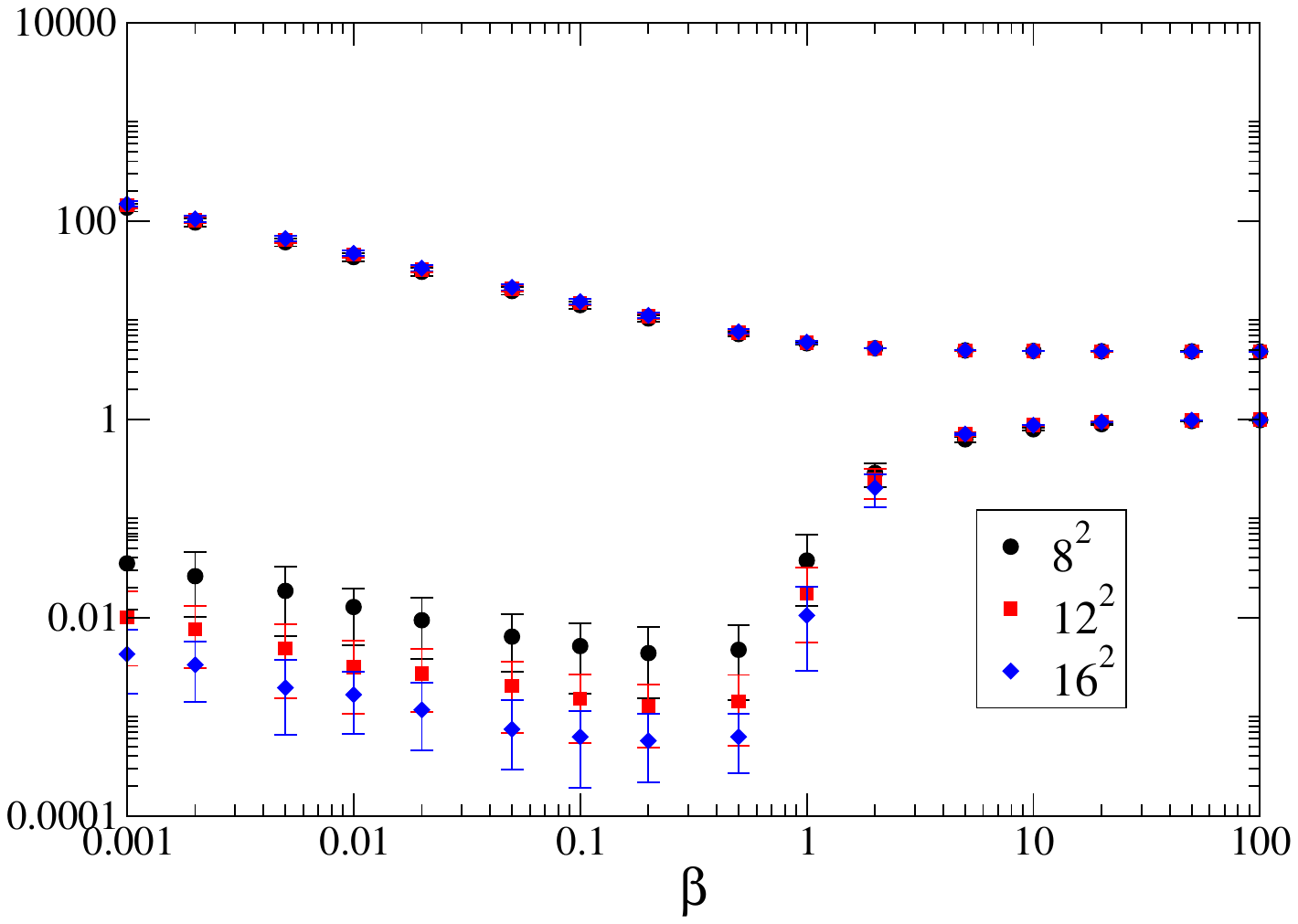}
\includegraphics[scale=0.22]{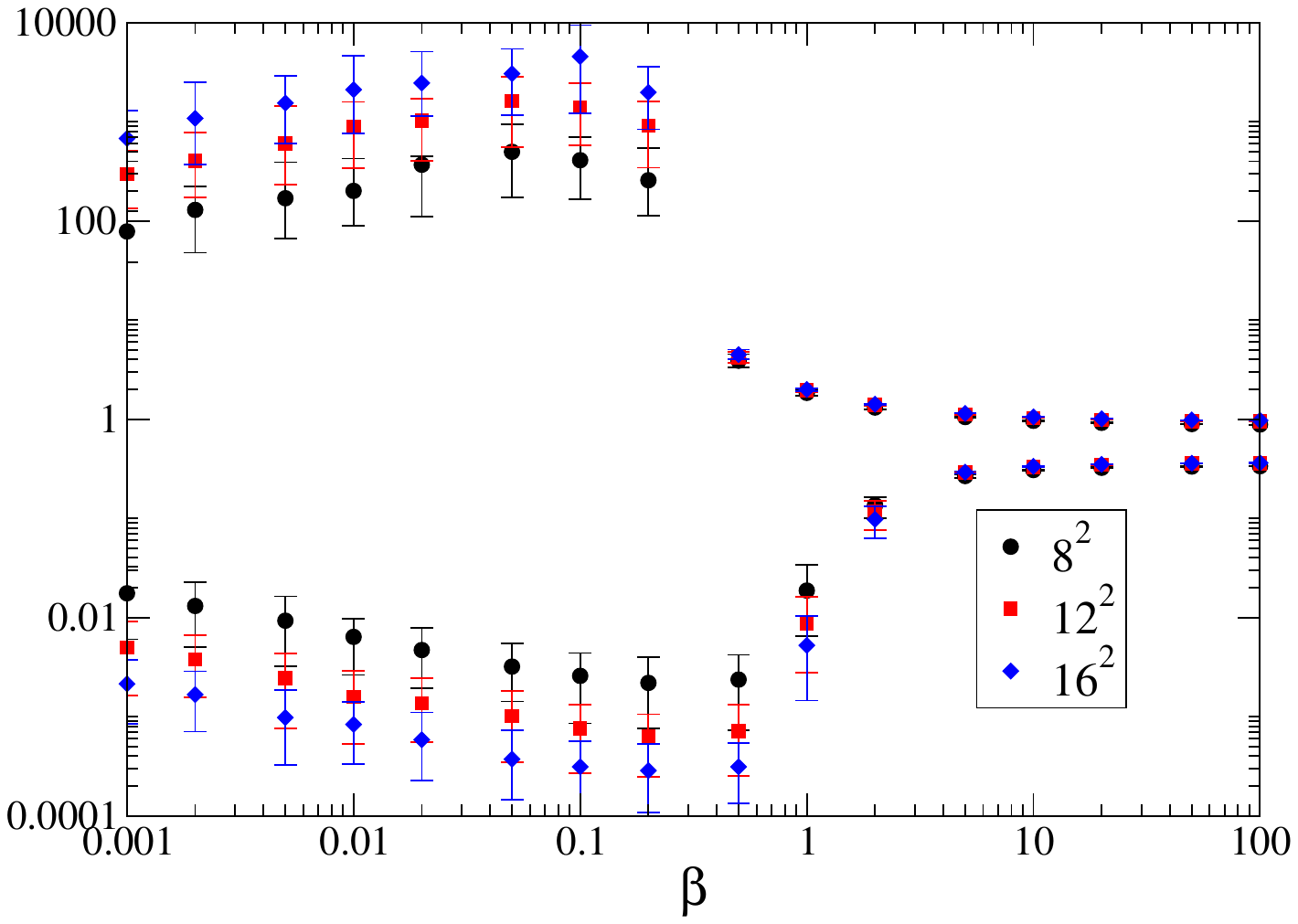}
\vspace{-0.5cm}
\end{center} \caption{Minimum and
maximum kernel eigenvalues for quenched auxiliary fields on different lattice
sizes (eg. $8^2$  in the key refers to lattice with $N_s=8$, $N_t=8$). 
Error bars show the first and ninth deciles.
Left panel: Wilson kernel. Right panel: Shamir kernel.}
\label{fig:qkminmax} \end{figure}

\begin{figure}[H] \begin{center}
\vspace{-0.5cm}
\includegraphics[scale=0.26]{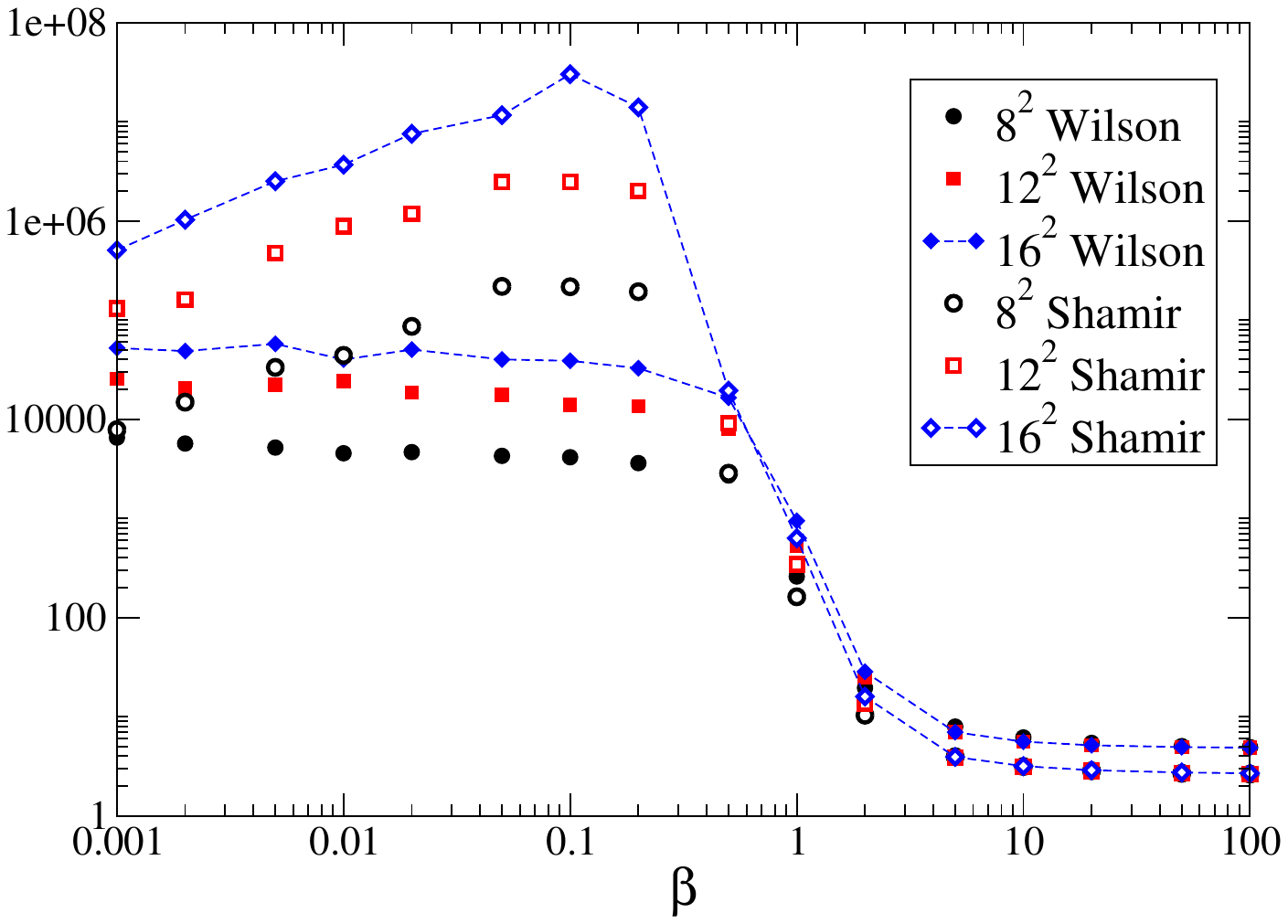}
\vspace{-0.5cm}
\end{center} \caption{Condition
number of the Shamir and Wilson kernel with quenched auxiliary fields and varying lattice
size.} \label{fig:qcond} 
\end{figure}

Figure \ref{fig:qkminmax} indicates the effect of mesh size $V$ on the eigenvalues,
this time with {\em quenched} auxiliary fields (ie. generated using solely the
gaussian measure $S=S_G[A]$ of eqn. (\ref{eqn:thirringaux}) rather than $S=S_F[\psi,\bar\psi,A]+S_G[A]$). 
The maximum eigenvalue is $V$-independent for the Wilson kernel. Noting that the critical region in this
case is in the vicinity of $\beta_c\approx0.7$, we find for the Shamir kernel, the maximum
eigenvalue is largely $V$-independent only on the symmetry unbroken
side. The smallest eigenvalues for both kernels are strongly dependent on
$V$ as $\beta$ moves into the strongly coupled region.  The lower
values reach a minimum somewhere around the critical region and then increase
again. Similarly for the strongly coupled side for Shamir kernel, the maximum
eigenvalue decreases again. However, the upper bound for the Wilson kernel is
monotonic. Whether the trends in these volume effects continue arbitrarily is
unclear from this data, although if continued it would suggest an unbounded
maximum eigenvalue for the Shamir case around the critical region. Figure
\ref{fig:qcond} shows the condition numbers. In the weakly coupled limit the
Shamir kernel has the lower condition number, and hence better numerical
properties, whereas moving towards the stronger coupling and through the
critical point the Wilson kernel has a much smaller condition number, although
the value declines again for the Shamir kernel.  

\subsection{Shamir Condensates}

\begin{figure}[H] \begin{center} 
\vspace{-0.5cm}
\includegraphics[scale=0.22]{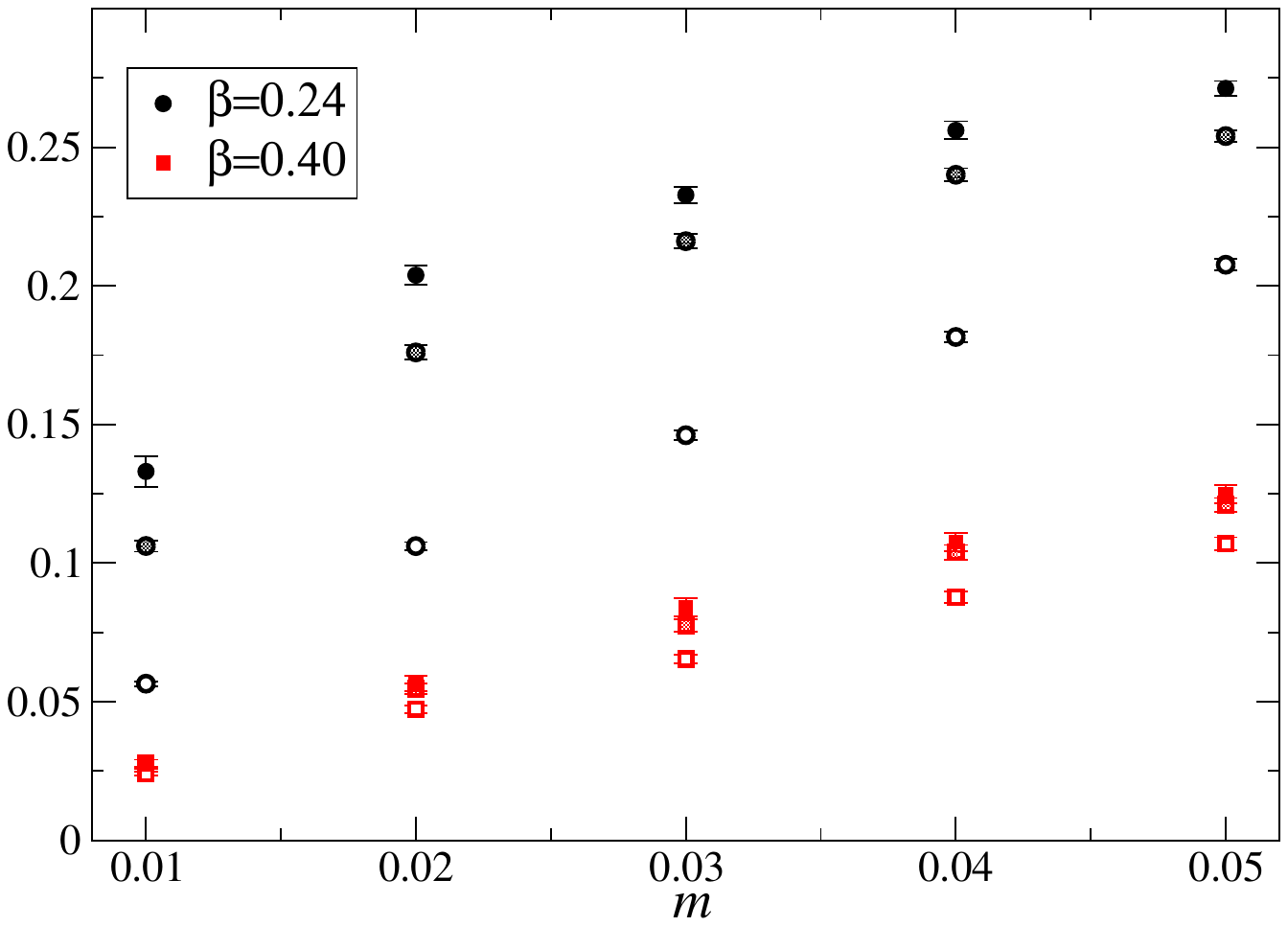}
\includegraphics[scale=0.22]{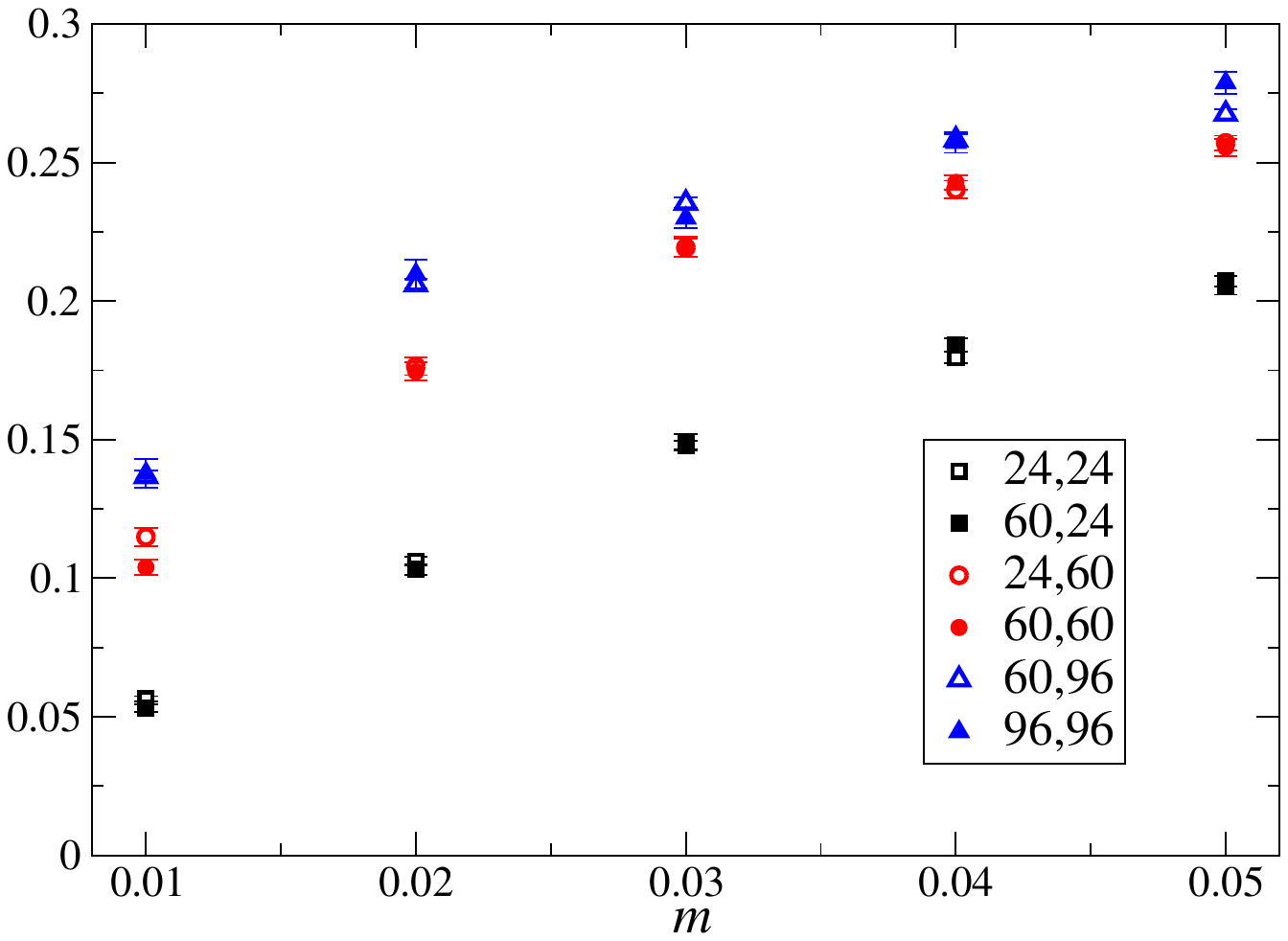}
\vspace{-0.5cm}
\end{center} \caption{Left:
Dynamic condensate plots with the Shamir kernel for
$\beta=0.24,0.40$ and $L_s=24$ (open symbols), 60 (shaded) and 96 (filled). Right: Partially
quenched Shamir condensates for $\beta=0.24$. X,Y in the legend denotes the auxiliary fields
were generated with $L_s={\rm X}$, and the measurements were taken with $L_s={\rm Y}$.
 At least 500 auxiliary field configurations were used in each measurement.
} \label{fig:ShamCond1} 
\end{figure}

Next we examine condensates evaluated with the Shamir kernel, similarly to
\cite{1} but using a smaller $12^2\times12$ mesh size rather than the
$16^2\times16$ meshes in those works. This is computationally beneficial not
only from the decreased mesh size, but also the expected decrease in eigenvalue
range of the kernel, which was noted in quenched simulations. Although
diminished, the $L_s$-limit challenges remain, and indeed the condensates
plotted as a function of mass $m$ in the left panel of fig. \ref{fig:ShamCond1} 
for $L_s=24,60,96$ do not suggest convergence in the strongly coupled region, even for $m=0.05$.

Following the observation that there may be no requirement for the auxiliary field
to be generated with such a stringent $L_s$ value, we look at partially quenched
condensates in the right panel of fig. \ref{fig:ShamCond1}, with a strong
coupling  $\beta=0.24$. As before, curves with sea and valence fermions calculated
with the same $L_s$ values are plotted, but now curves where the valence
fermions, and hence the condensates, are measured with a different $L_s$ value,
are also added. The results support the intuition that the condensate
measurements are controlled by the $L_s$ of the valence fermions, and using
lower $L_s$ values for the sea fermions has a limited impact on the results.
Given the high costs of dynamically generating the auxiliary fields, this
represents a significant potential in compuational cost cutting. Of course,
there is a certain inevitability that utilising different but sufficiently large
$L_s$ values for valence and sea fermions gives the same measurement results,
given the nature of $L_s$ convergence; for high enough values of $L_s$ the
limits of machine precision will eventually be reached. Given that measurements will only
ever be wanted to a certain accuracy, lower than machine precision, it seems quite reasonable that different
stages of the computational process may be performed with differing levels of
accuracy (which is what varying $L_s$ represents) and still achieve the final required accuracy.
 
The perspective of $L_s$ being a parameter of accuracy in the domain wall
formulation only makes sense when it is formally related to the overlap
operator. So-called non-bulk formulations in which the auxiliary (link) fields
$U$ applied in the Wilson operator $D_W[U]$ differ on each slice of the domain
wall extra dimension, in general have no formal relation to the overlap
operator, and hence there is no accuracy in a sign function to be considered. 

\begin{figure}[H]
\begin{center}
\vspace{-0.5cm}
\includegraphics[scale=0.22]{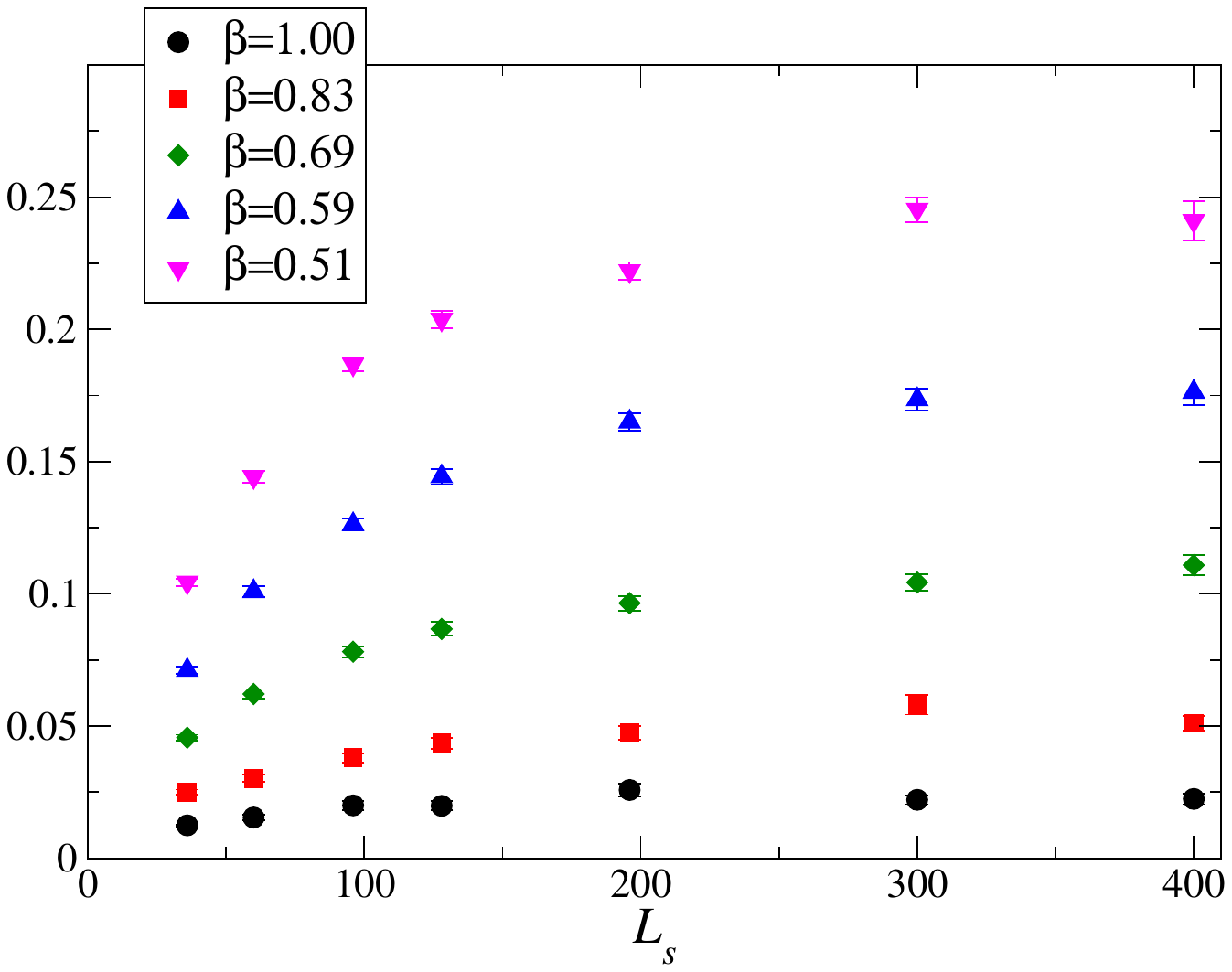}
\includegraphics[scale=0.22]{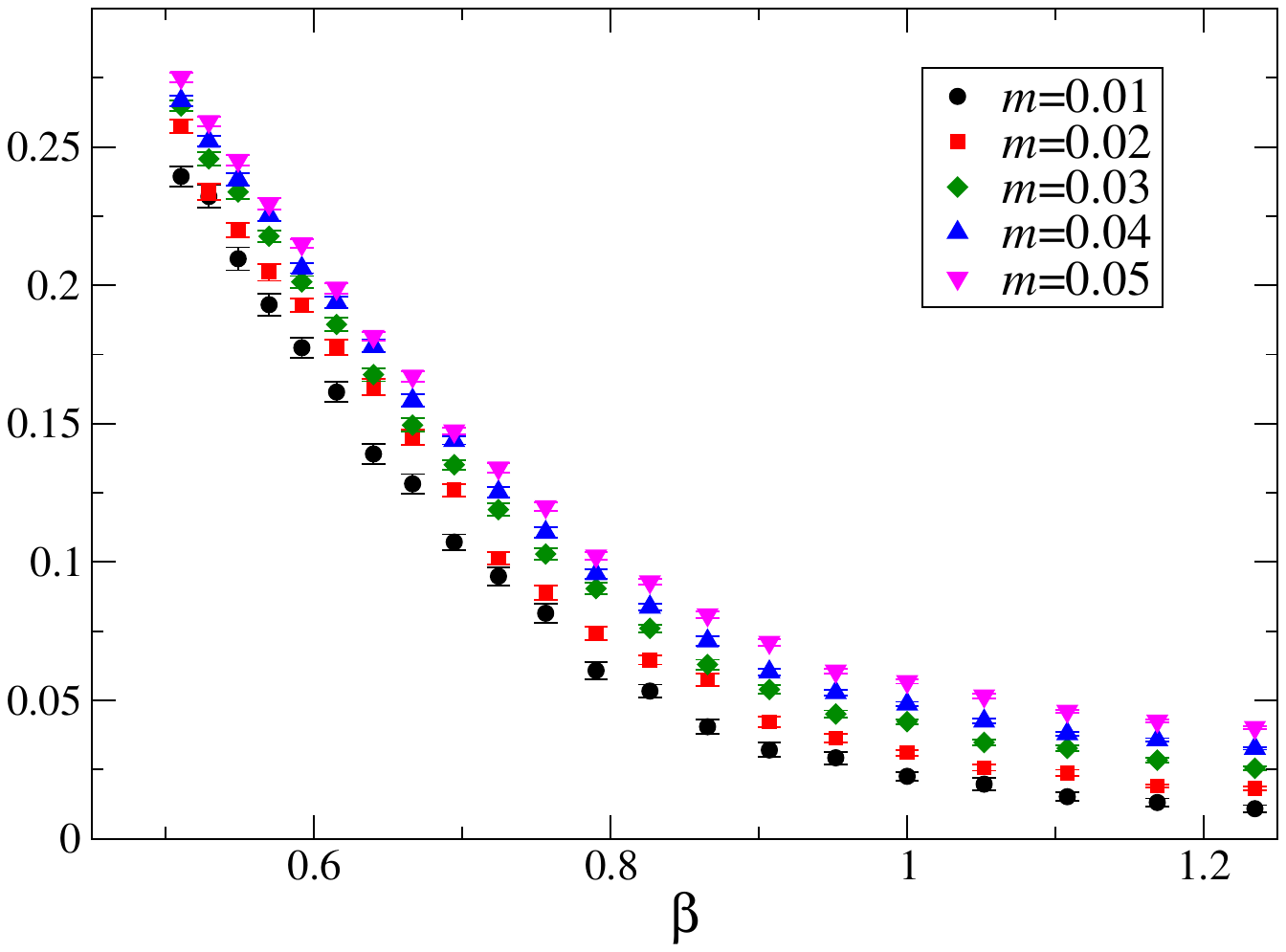}
\vspace{-0.5cm}
\end{center}
\caption{Quenched calculation of condensates using Shamir kernel. Left panel: vs
$L_s$ for different coupling strengths $\beta$ at $m=0.01$. 
Right panel: vs $\beta$ for different masses $m$ at $L_s$=300.}
\label{fig:QShamir}
\end{figure}

\begin{figure}[H]
\begin{center}
\vspace{-0.5cm}
\includegraphics[scale=0.22]{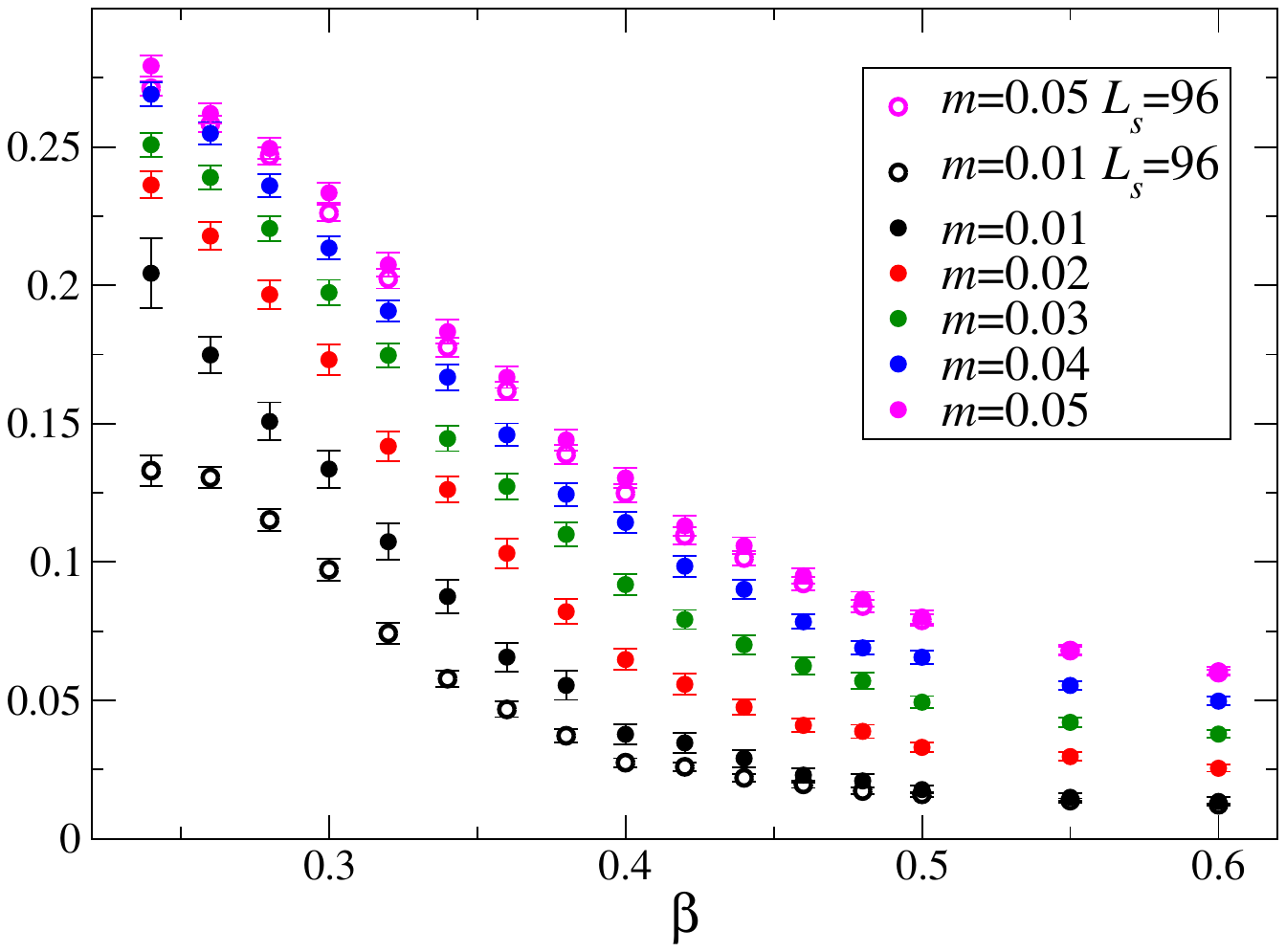}
\includegraphics[scale=0.22]{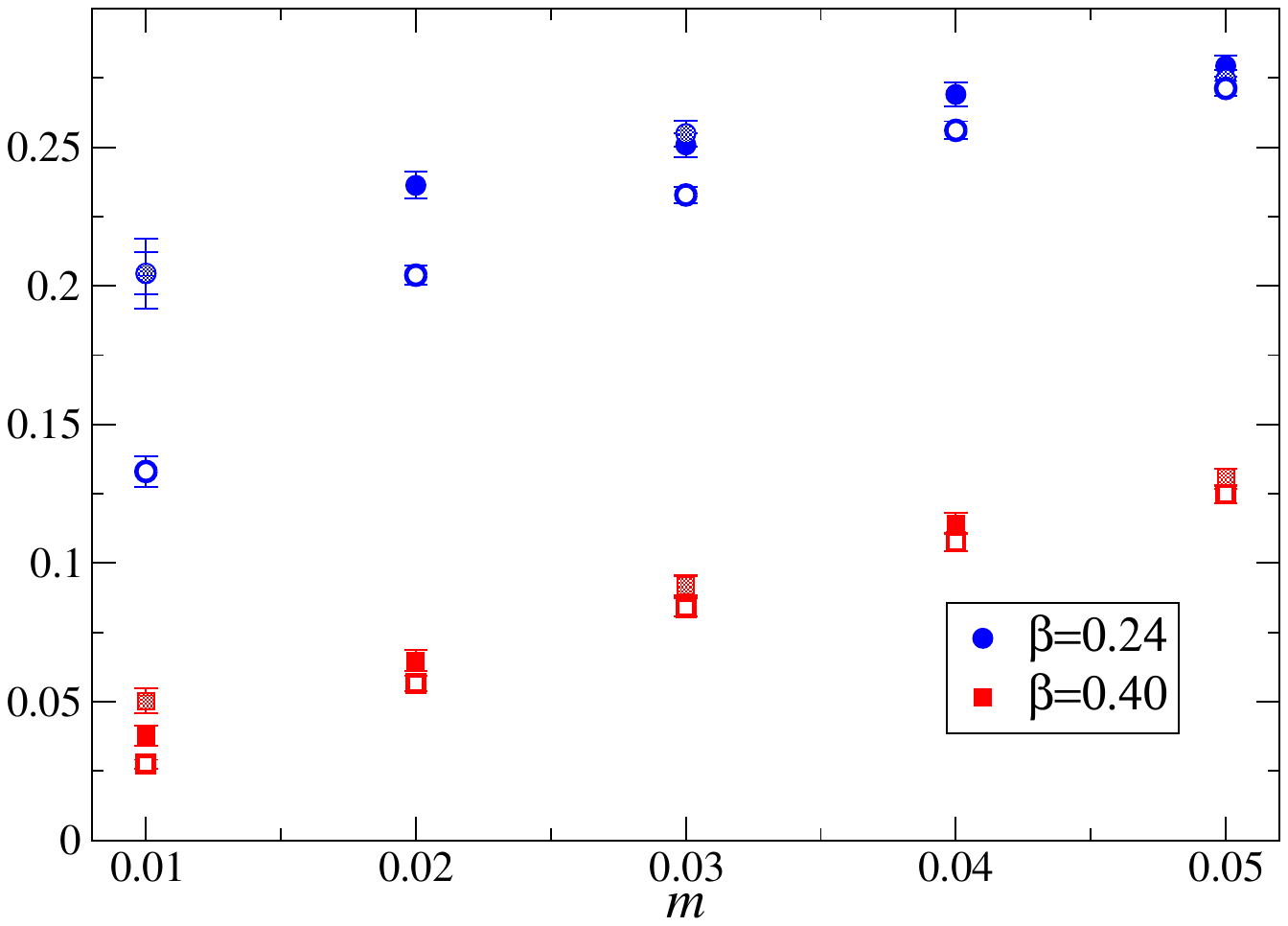}
\vspace{-0.5cm}
\end{center}
\caption{
Left panel: Partially quenched Shamir
condensates plotted vs $\beta$ using auxiliary fields generated with $L_s=96$. 
Measurements were taken with $L_s=300$ (filled) and $L_s=96$ (open). 
Right
panel: Partially quenched Shamir condensates measured with $L_s={\rm Y}$ plotted vs $m$. At least 50
configurations were used, generated with $L_s={\rm X}$, with (X,Y)=(96,300)
(filled), (60,300) (shaded) and (96,96) (open).} 
\label{fig:ShamCond3}
\end{figure}

We choose $L_s=300$ for the valence fermions in further {\em partially quenched} 
condensate measurements,
justified to a certain extent by the apparent convergence in quenched cases
shown in fig. \ref{fig:QShamir}. We use the auxiliary fields dynamically generated with
$L_s=96$. The left panel of fig. \ref{fig:ShamCond3} shows the results for 
$\langle\bar\psi\psi\rangle$ alongside
those measured with $L_s=96$. They show that the convergence has not set in
by $L_s=96$
even at relatively weak coupling. By contrast, supposing that the
$L_s=300$ is sufficient for capturing U(2)-invariant dynamics, then the right hand panel demonstrates that
the sea fermions were adequately modelled, and that $L_s=60$ or even lower would
have been sufficient for the sea fermions. Condensate instances were calcuated
every 10 trajectories of the hamiltonian step, with at least 2000 trajectories taken.

\subsection{Wilson Condensates}

\begin{figure}[H] 
\begin{center}
\vspace{-1.2cm}
\includegraphics[scale=0.32]{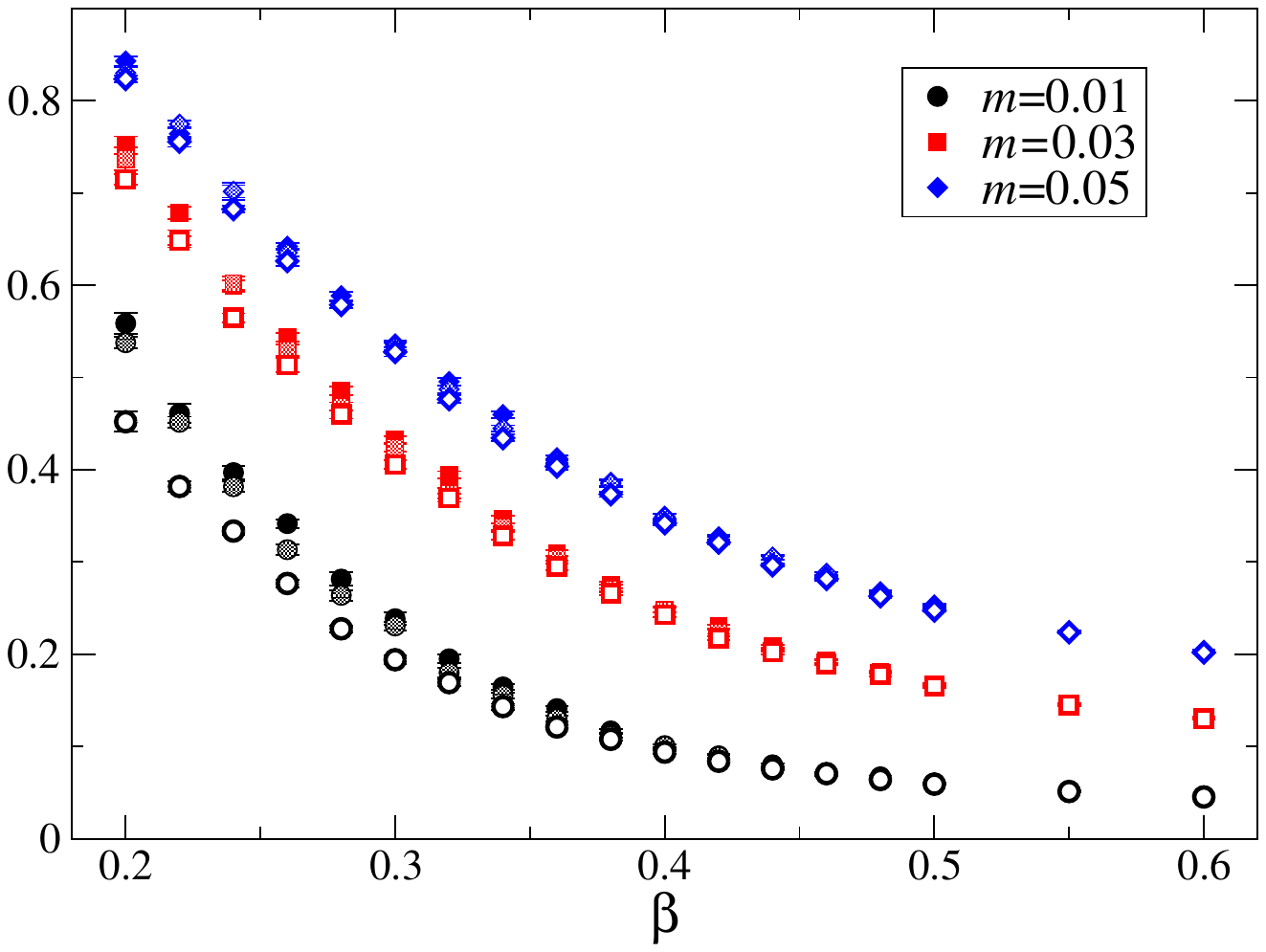}
\vspace{-0.5cm}
\end{center} 
\caption{Dynamic Condensate with Wilson kernel for various $m$, 
using HT formulation with $L_s=36$ (flled), 28 (shaded), 20 (open).} 
\label{fig:DynWCond1} 
\vspace{-2.0cm}
\end{figure}

Figure \ref{fig:DynWCond1} shows condensates generated with the Wilson kernel,
with instances taken every
5 trajectories over at least 1500 trajectories, using the hyperbolic tangent (HT)
approximation formulation with various $L_s$. 
Eyeballing the results suggests that the $m=0.05$ case seems to be well
$L_s$-converged
already at $L_s=20$, and it appears that it may be close to
satisfactory convergence for $L_s=28,36$ for $m=0.01$ and $m=0.03$.

\begin{figure}[H]
\begin{center}
\vspace{-0.5cm}
\includegraphics[scale=0.22]{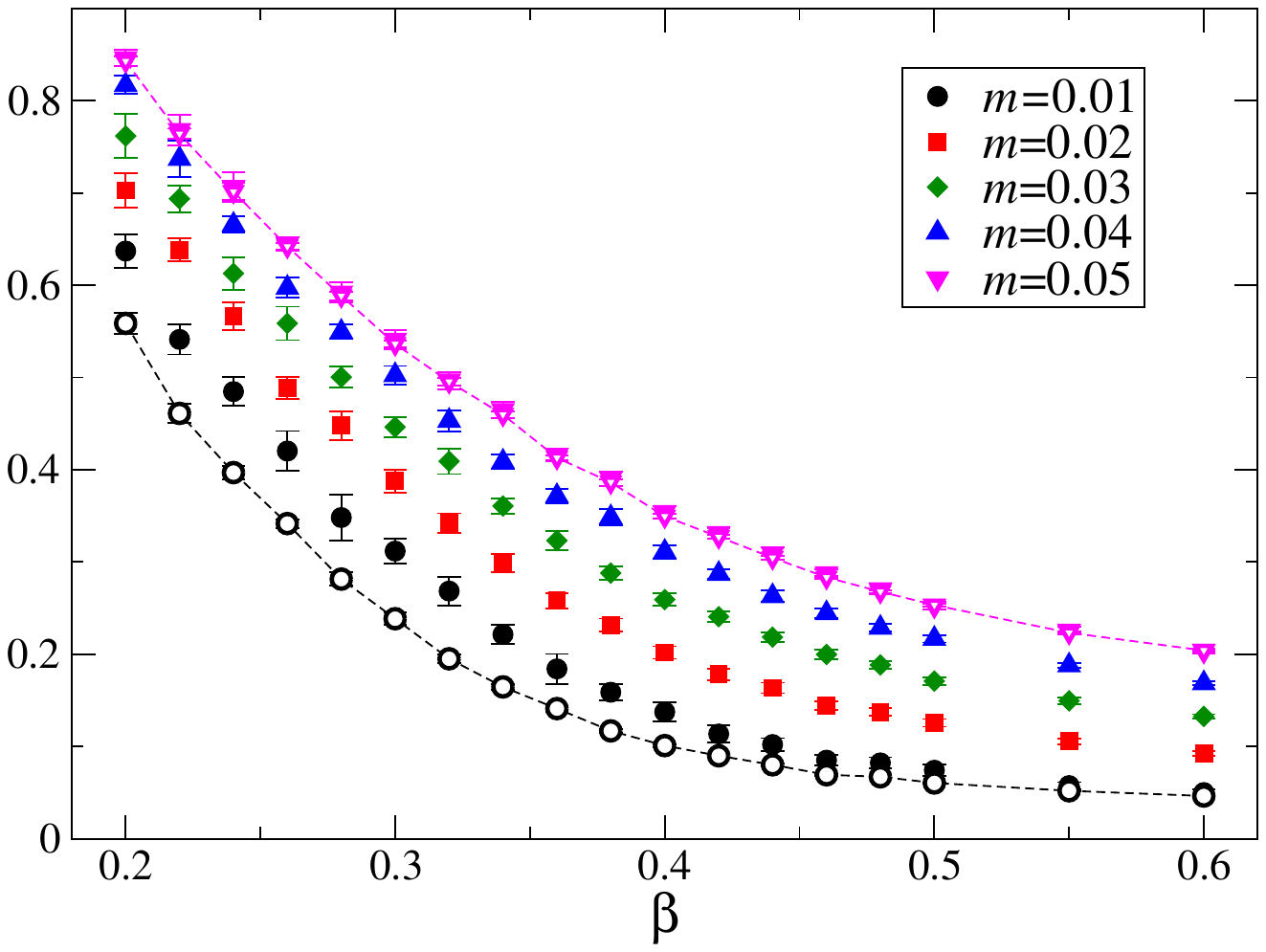}
\includegraphics[scale=0.22]{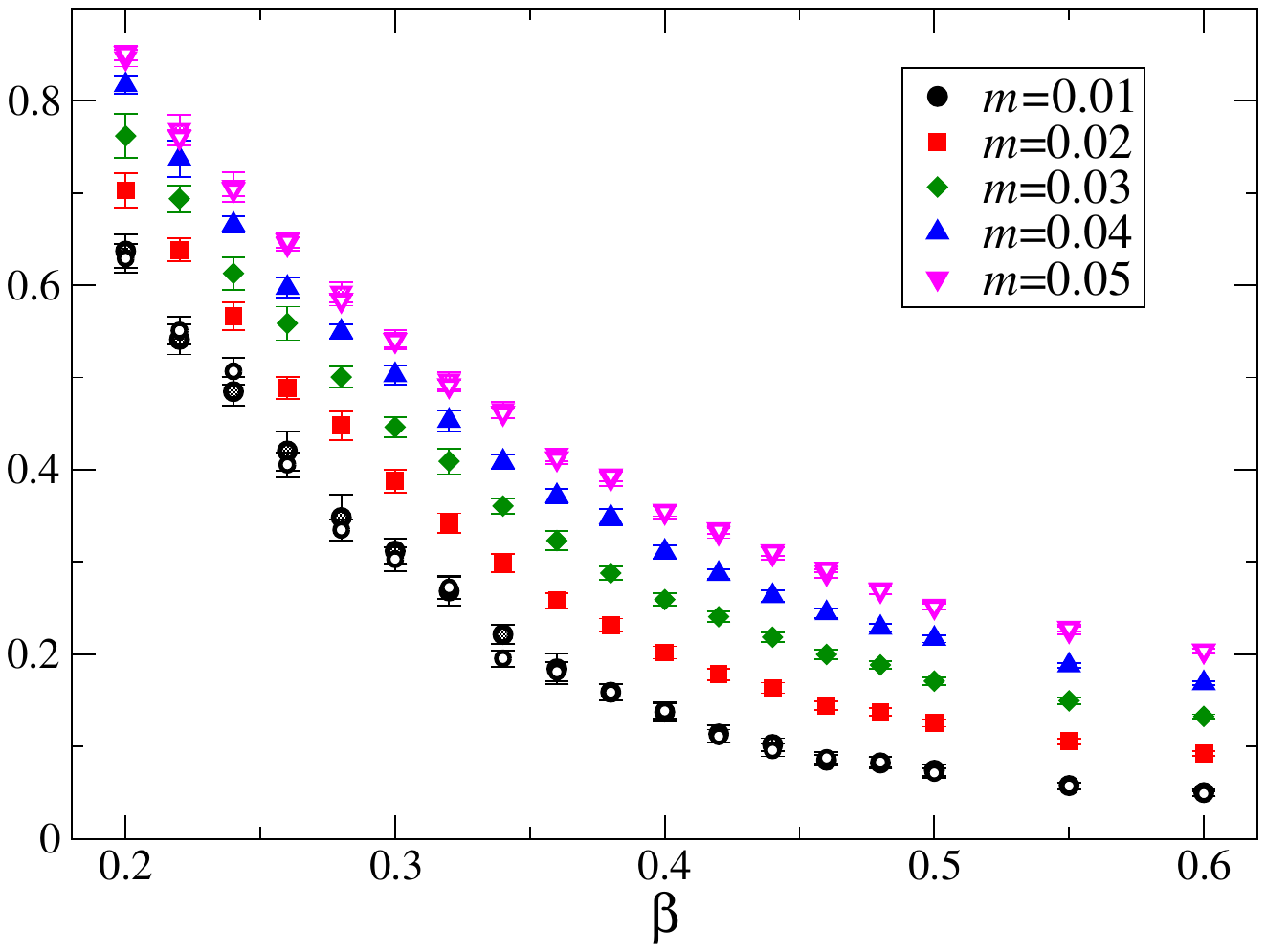}
\vspace{-0.5cm}
\end{center}
\caption{Dynamic Condensate with Wilson kernel. The auxiliary fields were dynamically
generated with HT and $L_s=36$. Left panel: Measurements with Zolotarev
($L_s=24$, Range=$[0.001,20]$ (filled symbols)) and HT ($L_s=36$, open) 
formulations at different $m$.
Right panel: Measurements with Zolotarev (Range=$[0.001,20]$: $L_s=18$ (open), 24 (shaded);
Range=$[0.0005,20]$: $L_s=30$ (filled))} \label{fig:DynWCond2} 
\end{figure}

This is misleading, however. The left panel of Figure \ref{fig:DynWCond2} shows
measurements using the Zolotarev formulation for the valence fermions with
$L_s=24$ and a range $[0.001,10]$, against two different mass plots of the HT
formulation. While we again see that at $m=0.05$ the condensate appears
$L_s$-converged,
we also see that the HT formulation is not yet converged for $m=0.01$. We argue
that similarly to the Shamir case, we do not require the $L_s$-convegence in the
generation of the auxiliary field to match that in the measurements, and hence
proceed using  HT-generated auxiliary fields. We also want to check that
the Zolotarev range and $L_s$ value is sufficient. The right panel of Figure
\ref{fig:DynWCond2} shows the condensates measured (still using the $L_s=36$
generated auxiliary fields) with Zolotarev range $[0.001,20]$ for $L_s=18$ and
$L_s=24$, expanded to $[0.0005,20]$ for $L_s=30$. Based on the errors of the
scalar sign function approximation we might have expected more stringent
conditions to be necessary, but it appears that reasonable $L_s$-convergence is
already achieved by $L_s=18$.

\subsection{Equation of State}

\begin{figure}[H]
\begin{center}
\includegraphics[scale=0.35]{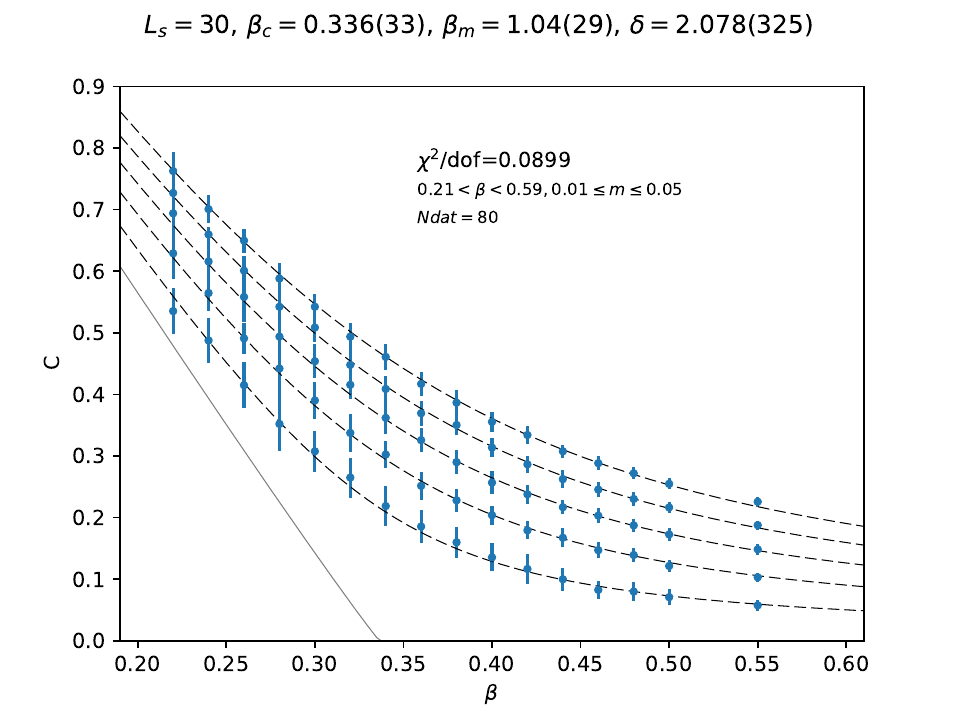}
\includegraphics[scale=0.35]{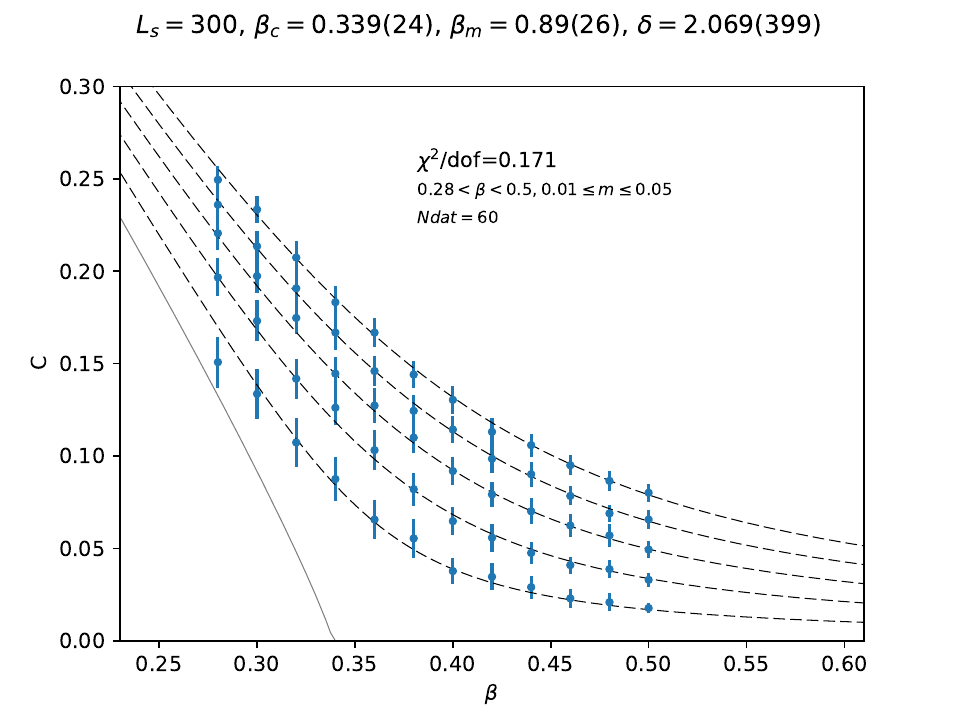}
\end{center}
\caption{Equation of State fits. Left panel:  Wilson formulation. Auxiliary fields generated with HT and $L_s=36$, measurements with Zolotarev and $L_s=30$. Right panel: Shamir formulation. Auxiliary fields generated with HT and $L_s=96$, measurements with Zolotarev and $L_s=300$.}
\label{fig:EOS}
\end{figure}

Finally we want to calculate an equation of state \cite{5} from the condensate
measurements, given by eqn. (\ref{eqn:EoS}), and estimate the critical exponents.
For Dirac operators which have the same physical content, we would expect the
fitted exponents to be the same: this is a manifestaton of the universality
expected at a critical point regardless of the microscopic details of the
regularisation procedure employed.  Hence the choice of Wilson or Shamir kernel
should yield the same critical exponents. Fig. \ref{fig:EOS} shows two such fits
with the corresponding exponents set out in table \ref{tab:EOS}. As has been
noted before \cite{1}, the values are sensitive to the data set chosen to be
fitted. Nevertheless we find that the results are in plausible agreement. 

\begin{table}[ht]
\tbl{Equation of state critical exponents found with partially quenched Shamir HT kernel for different $\beta$ data range windows. Mass range is $[0.01,0.05]$.}
{\begin{tabular}{@{}ccccccc@{}} \toprule
\hline
  & $\beta$ & $\beta_c$ & $\beta_m$ & $\delta$ & $\eta$ & $\nu$ \\
 \hline
 Wilson  & 0.22-0.55 & 0.336(33) & 1.04(29) & 2.078(325) & 0.95(15) & 1.1(3)  \\ 
 Shamir & 0.28-0.50 &  0.339(24) & 0.89(26)   & 2.069(399) & 0.96(18) & 0.91(28) \\
 \hline
\end{tabular} \label{tab:EOS}}
\end{table}

The critical exponents are consistent between the Shamir and Wilson
formulations, although for $\beta_m$ this within fairly large error margins. 
Of course, this
data only hints at the similarity of the results, and more and better data is
required. 
However, significant differences are found from the values given in \cite{18,25}.
The exponents found with a Shamir kernel on $16^2\times16$ mesh were
$\beta_c=0.320(5)$, $\beta_m=0.320(5)$, $\delta=4.17(5)$, corresponding to
$\nu=0.55(1)$ and $\eta=0.16(1)$. We attribute the difference to the lack of
$L_s$-convergence in earlier work. Further comparison with a staggered
formulation may be considered which gave~\cite{5}  $\beta_m=0.57(2)$, $\delta=2.75(9)$,
corresponding to $\nu=0.60(4)$ and $\eta=0.71(3)$. Differences here are thought
to stem from formulational differences.

\section{Summary}

We have made progress in overcoming the challenging $L_s$ limit of the overlap
operator in the context of the planar Thirring model at a critical point,
yielding distinct results from similar earlier enterprises, albeit on a
relatively small mesh size. 

A number of different aspects of the Dirac operators and their implementations
were investigated. Eigenvalue ranges of the overlap kernel were explored, and a
key finding is that the non-compact link fields leads to kernel eigenvalues apparently unbounded
from above, whereas these are clearly
bounded with compact link fields. This has very significant (detrimental)
implications on the computational difficulty of inverting the Dirac operator.
Further, the Shamir formulation appears to become more challenging as the mesh
gets larger in a way that the Wilson kernel doesn't. That is, that the largest
kernel eigenvalue increases with mesh size around and beyond the critical
coupling strength.

A key finding, or observation, to carry to future work, was that the level of
accuracy corresponding to the large $L_s$ limit required in the measurement of
the condensate, is not required in the generation of the auxiliary fields. Given
that the bulk of computational effort in dynamic simulations is in the
generation of the auxiliary fields this has the potential to save significant
computational cost. Further, rather than being a tradeoff with the Monte Carlo
acceptance rate, $L_s$ reduction improves the acceptance rate.

The (expected) superiority of the Zolotarev approximation was demonstrated to be
necessary to achieve $L_s$-convergence of the measurements with a modest value
of $L_s$, although a modest $L_s$ value with the HT approximation appeared to be
sufficient for the auxiliary field generation.

We calculated an equation of state and critical exponents for both Wilson and
Shamir kernels, using low accuracy for the generation of the auxiliary fields
and high accuracy for the measurements. Although better statistics would be
desirable, the results are consistent between the formulations, as we would
hope. This was particularly good news as lower $L_s$-range preliminary work had
hinted that consistency may not have been found. This would not have been
possible without partial quenching.

\section*{Acknowledgments}
This work used the DiRAC Data Intensive service (CSD3) at the University of
Cambridge, managed by the University of Cambridge University Information
Services on behalf of the STFC DiRAC HPC Facility (www.dirac.ac.uk). The DiRAC
component of CSD3 at Cambridge was funded by BEIS, UKRI and STFC capital funding
and STFC operations grants. DiRAC is part of the UKRI Digital Research
Infrastructure. Further work was performed on the Sunbird facility of
Supercomputing Wales. The work of JW was supported by an EPSRC studentship, and
of SH by STFC Consolidated Grant ST/ST000813/1.


\end{document}